\documentclass[usenatbib,useAMS,a4paper]{mn2e}

\usepackage{txfonts}
\usepackage{epsfig}
\newcommand{\aap}{A\&A}

\newcommand{\mnras}{MNRAS}
\newcommand{\nat}{Nature}
\newcommand{\apj}{ApJ}
\newcommand{\apjl}{ApJL}
\newcommand{\apjs}{ApS}
\newcommand{\aj}{AJ}

\title [Spectra of photoevaporating discs]{Theoretical spectra of photoevaporating protoplanetary discs: \\
An atlas of atomic and low-ionisation emission lines}
\author[]{Barbara Ercolano$^{1,2,3}$ and James E. Owen$^{2}$\\
$^1$Department of Physics and Astronomy, University College London, WC1E 6BT, UK\\
$^2$Institute of Astronomy, Madingley Rd, Cambridge, CB3 0HA, UK \\
$^3$Current Affiliation: School of Physics, University of Exeter, Stocker Road, Exeter EX4 4QL}

\date{Submitted:}

\pagerange{\pageref{firstpage}--\pageref{lastpage}} \pubyear{2003}

\begin{document}

\label{firstpage}
\maketitle

\begin{abstract}

We present a calculation of the atomic and low-ionisation emission
line spectra of photoevaporating protoplanetary discs. Line
luminosities and profiles are obtained from detailed photoionisation
calculations of the disc and wind structures surrounding young active
solar-type stars. The disc and wind density and velocity fields were
obtained from the recently developed radiation-hydrodynamic models of
Owen et al., that include stellar X-ray and EUV irradiation of protoplanetary 
discs at various stages of clearing, from primordial sources to inner
hole sources of various hole sizes. 

Our models compare favourably with currently available observations, 
lending support to an X-ray driven photoevaporation model for disc
dispersal. In particular, we find that X-rays drive a warm,
predominantly neutral flow where the OI~6300{\AA} line can be produced
by neutral hydrogen collisional excitation. Our  
models can, for the first time, provide a very good match to both
luminosities and profiles of 
the low-velocity component of the OI~6300{\AA} line and other
forbidden lines observed by Hartigan 
et al., which covered a large sample of T-Tauri stars. 

We find that the OI~6300{\AA} and the
NeII~12.8$\mu$m lines are predominantly produced in the X-ray-driven
wind and thus appear blue-shifted by a few km/s for some of the
systems when observed at non-edge-on inclinations. We note however that blue-shifts
are only produced under certain conditions: X-ray luminosity,
spectral shape and inner hole size all affect the location of the
emitting region and the physical conditions in the wind. We caution
therefore that while a blueshifted line is a tell-tale sign of an
outflow, the lack of a blueshift should not be necessarily interpreted
as a lack of outflow. 
Comparison of
spectrally resolved observations of multiple emission lines with
detailed model sets, like the one presented here, should provide useful diagnostics
of the clearing of gaseous discs. 

%We note that our models underpredict the luminosities of IR hydrogen
%recombination lines by one or two order of magnitudes, implying a non
%disc/wind origin for this line, which is probably produced in very
%dense plasma close in to the star. 

\end{abstract}


\section{Introduction}

The circumstellar environments of newly-born low to intermediate
mass stars have recently been the object of much interest for their
potential of hosting planetary systems. These (protoplanetary) discs of
dust and gas hold the reservoir of material from which planets may
form and provide the medium through which newly formed planets and
planetesimals migrate towards or away from their parent star. For this
reason no theory of planet formation 
can be complete without an understanding of how protoplanetary discs
evolve and finally disperse. 

The dispersal of protoplanetary discs has been studied
observationally, through statistics of disc fractions in young
clusters (e.g. Haisch et al 2001; Hernandez et al 2008; Mamajek 2009)
and on an object-by-object basis through observations of individual
so-called transition discs (e.g. Calvet et al 2002), 
i.e. discs that appear to be optically thin in the dust in their inner
regions, which is generally interpreted as evidence of partial
(inside-out) clearing. Whether these discs with inner-holes
  really represent a stage in the evolution/dispersal of
  protoplanetary discs is currently still uncertain.
Currently, the  
consensus in the literature is that the dispersal of discs around
T-Tauri stars is characterised by a rather fast transition timescale,
evidenced by the fact that very few objects (typically 10\%) are
caught in this transition phase (Strom et al. 1989; Strutskie et
al. 1990; Ercolano, Clarke \& Robitaille 2009; Kim et al 2009; Luhman
et al 2010; 
Muzerolle et al 2010)\footnote{There have been 
claims of longer transition timescales for M-stars or older clusters
  (Sicilia-Aguilar et al. 2008; Currie et al 2009), these results,
however, are still object of debate in the literature (Ercolano et al
2009, Muzerolle et al. 2009, Luhman et al 2010)}. 

On the basis of these observational results, theoretical efforts have
thus focused on identifying a mechanism to provide a two-timescale
dispersal phenomenon, i.e. a mechanism that 
would allow protoplanetary discs to be detected as optically thick (in
the dust) for a few million years and then cause them to disperse very
rapidly in about a tenth of their total lifetime, i.e. a timescale much
shorter than what would be expected by viscous evolution alone (Hartmann et al. 1998, Clarke et al. 2001). 

The proposed mechanisms include planet formation itself
(e.g. Armitage \& Hansen 1999), grain growth (Dullemond \& Dominik 2005), photophoresis (Krauss et al 2007), MRI-driven winds (Suzuki \& Inutsuka 2009), and 
photoevaporation (Clarke et al 2001; Alexander et al 2006, Richling \& Yorke 2000, Ercolano et al 2008b, 2009a, Gorti \& Hollenbach 2009; Owen et al. 2010a).
Understanding which (or what combination) of the proposed mechanisms
dominates dispersal is key to understanding disc evolution and
thus provides important constraints on planet formation theories by setting
the timescale over which planets may form and migrate. 

Photoevaporation by soft X-ray radiation (100~eV $<$ E $<$ 1keV) has been shown to drive powerful 
winds (Ercolano et al. 2008b, 2009a, Gorti \& Hollenbach 2009), able to disperse the discs %added comma after winds
rapidly once the accretion rates have 
fallen to values lower than the photoevaporation rates
($\sim$10$^{-8}$M$_{\odot}$/yr for X-ray luminosities of
2$\times$10$^{30}$erg/s; Owen et al. 2010a; from now on O10). The %removed the word wind to make the sentance flow better
mass-loss rates are roughly proportional to the impinging X-ray  %added hyphen in mass-loss
luminosity (Ercolano 2008b), meaning that the expected mass-loss rates (and hence accretion %added reference to where Lx vs Mdot is discovered changed wind to mass-loss
rates) for solar type stars span over the two-three decade range of X-ray%added three
luminosities(e.g. Preibisch et al. 2005; G\"udel et
al. 2007).%add refence here  
 The recent observation of shorter disc lifetimes in low-metallicity
 environments 
(Yasui et al 2009) has been interpreted as further evidence for
X-ray photoevaporation acting as the main disc dispersal mechanism in
those clusters (Ercolano \& Clarke 2009). 

While encouraging, the evidence in favour of photoevaporation produced
so far consists of statistics derived from observations %added comma and replace second of with derived from 
of disc dispersal based on the evolution of their dust
component. However, dust represents only a percent or so of the total
disc mass and the 
evolution of the dust and gas component of a disc do not necessarily
go hand in hand (e.g. Alexander \& Armitage 2007; Takeuchi, Clarke \&
Lin 2004). In the last few years, observations of %added comma 
emission lines from warm atomic and low ionisation gas in the
inner region of discs are starting to provide some helpful clues
of the physical conditions of the gaseous component in these
planet-forming regions (Hartigan et al 1995;Herczeg et al 2007;
Pascucci et al 2007; Razka et al 2007; Espaillat et al 1995;
Najita et al 2009; Pascucci \& Sterzik 2009; Flaccomio et al 2009;
G\"udel et al 2010, submitted). Line emission from low   
ionisation and atomic species is
 of particular interest, as it is likely to trace material that is located
in the warm upper layers of the disc atmosphere or in the flow
itself, allowing the response of the gas to irradiation from the
central star to be probed and understood.
In particular Pascucci \& Sterzik (2009) reported evidence of a
blue-shifted NeII fine structure line at 12.8$\mu$m, the blue-shift is
consistent with the line being formed in a slow moving (a few km/s)
disc wind (Alexander 2008). Earlier spectrally resolved observations
of optical forbidden lines in T-Tauri stars (Hartigan et al., 1995,
from now on HEG95),
including OI 6300~{\AA}, also revealed blue-shifted low velocity
components, interpreted as evidence for a disc wind. The
interpretation of these atomic and low-ionisation emission line
profiles is key to understand the kinematic structure of the
warm upper disc layers and its wind. In this paper we provide
the tools to exploit currently available and future observations using
the framework of the X-ray driven photoevaporation model of O10 (further discussed in Section~\ref{s:m}). 

Furthermore, with the {\it Herschel Space Telescope} and, later, {\it James Webb
  Space Telescope}, committed to studying planet formation through the
characterisation of discs, it is both timely and important to provide a solid
theoretical framework for the interpretation of these new%changed the to these
observations. In this context this paper has two main aims: (i) To %capatilised To 
provide an atlas of atomic and low ionisation emission line
intensities and profiles predicted for the photoevaporating
protoplanetary discs of O10 at various stages in 
their evolution; (ii) To compare the %capatilised To
predictions of O10's X-ray driven photoevaporation model with
currently available observations and search for the `smoking gun' of
photoevaporation. We 
therefore list line intensities for the 
brightest lines and calculate line profiles for some of the more
promising wind/disc tracers. We show results for discs at different
stages of clearing,  i.e. photoevaporating primordial discs where an
inner hole is yet to form and systems in a more advanced clearing
stage with inner holes (in dust and gas) of increasing
radius. We note 
that, while several authors  
have presented theoretical calculations of emission line luminosities
and profiles from T-Tauri discs (e.g. Glassgold et al 2007,
Meijerink et al. 2008, Ercolano et al 2008b, Gorti \&
Hollenbach 2008, Hollenbach \& Gorti 2009, Schisano, Ercolano \& G\"udel 2010), this is 
the first time that a fully self-consistent radiation-hydrodynamical
and photoionisation calculation is attempted, which provides us with
the three-dimensional distribution of gas velocities and emissivities
in the heated layers of the disc and its wind. 

A comparison of the calculations presented in this paper with
available observations suggests that X-ray heated discs with
hydrodynamically escaping winds produce emission line intensities and
profiles that are 
consistent with those that are currently being observed. It is hoped
that the results presented here may be used for the interpretation of
future observational data which will be crucial to further test
the validity of the X-ray photoevaporation scenario as a dominant disc
clearing mechanism. 

The paper is organised as followed. Section~2 briefly describes the
radiation-hydrodynamic disc models and lists the input parameters. Sections~3 presents the results,
including the emission line intensities and the predicted
profiles for selected transitions. 
Sections~4 and~5 deal with a comparison of our results with currently
available data. The main points of the paper are briefly summarised in Section~6. 

\section{The radiation-hydrodynamics models}
\label{s:m}

Energetic radiation from a young stellar objects (YSOs) ionises and
heats the gas in the surrounding protoplanetary disc.  The
gas expands as a result of the extra heating and some regions of
the disc atmosphere may become unbound, giving % removed for given conditions
rise to a photoevaporative wind from the disc surface. A calculation
of accurate mass loss rates and of the density and geometry of the
flow requires the simultaneous solution of the two-dimensional radiative
transfer and the hydrodynamics problem. Owen et al. (2010a, O10) recently
presented the first fully self-consistent radiation hydrodynamics
calculations of an X-ray plus EUV photoevaporated protoplanetary
discs. Ercolano et al (2009) and O10 found irradiation by soft X-rays
to be most efficient at producing a photoevaporative disc wind with 
mass loss rates of $\sim10^{-8}$M$_{\odot}$/yr for X-ray luminosities
of 10$^{30}$ erg/sec. When coupled with viscous evolution,  
these wind rates are sufficiently large to produce an inner hole in
the disc, when the accretion rate drops significantly (a factor
$\sim$10) below the wind mass-loss rates. The remaining material is then
dispersed in the final 15-20\% of the disc lifetime.
  In this paper we refer to EUV 
  radiation as having energies between 13.6eV and 100~eV, soft X-ray
  radiation between 100~eV and 1~keV, and hard X-ray radiation for
  energies larger than 1~keV. The total 'X-ray' luminosity is defined
  between 0.1~keV and 10~keV, which is similar to the bandpass quoted
  by G\"uedel et al 2009 (0.3-10~keV).

Two-dimensional hydrodynamic calculations were performed using
a parameterisation of the gas temperature based on the results of
three-dimensional radiative transfer and photoionisation
calculations. Radiative equilibrium was 
assumed and verified a posteriori by ensuring that the thermal %changed cooling to thermal
timescale is shorter than the flow timescale.  %changed faster to shorter
The hydrodynamic calculations were carried out using a modified
version of the {\sc zeus-2d} code (Stone \& Norman 1992a,b; Stone et
al 1992) which includes heating from the X-ray and EUV radiation from a
central YSO. The radiative transfer and photoionisation calculations
were carried out using the 3D {\sc mocassin} code (Ercolano et
al. 2003, 2005, 2008a) modified according to Ercolano et al. (2008b,
2009a). The atomic database included opacity data from
Verner et al. (1993) and Verner \& Yakovlev (1995), 
energy levels, collision strengths and transition probabilities from
Version 5.2 of the CHIANTI database (Landi et al. 2006, and references
therein) and the improved hydrogen and helium free-bound continuous
emission data of Ercolano \& Storey (2006).

The models presented here are appropriate for a
 0.7~M$_{\odot}$ star; models for
lower/higher mass stars are currently being  
developed and will be presented in a future paper. We vary the X-ray luminosity to span the observed range for
classical T-Tauri stars (cTTs) of this mass ( 2$\times$10$^{28} \leq $L$_X \leq $
2$\times$10$^{30}$ erg/sec; e.g. Preibisch et al. 2005; G\"udel et
al. 2007) and consider discs at different stages of dispersal, from
discs extending all the way into their dust destruction radius
('primordial discs') to discs with cleared inner holes of varying
sizes (R$_{in}$ = 8.3~AU, 14.2~AU and 30.5~AU, models A, D and G in O10, respectively). 
The ionising spectra used were calculated by Ercolano et al (2009a),
using the plasma code of Kashyap \& Drake (2000) from an emission
measure distribution based on that derived for RS CVn type binaries by
Sanz-Forcada et al. (2002), which peaks at 10$^4$~K and fits to
Chandra spectra of T-Tauri stars by Maggio et al. (2007), which peaks
at around 10$^{7.5}$~K. For the remainder of this paper we will refer
to the models by their X-ray luminosity
(0.1~keV~$<$~h\,$\nu$~$<$~10~keV), we note, however, that the
irradiating spectrum has a significant EUV component
(13.6~eV~$<$~h\,$\nu$~$<$~0.1~keV) with L$_{EUV}$~=~L$_{X}$ (see
Ercolano et al 2009a for further discussion).

The following elemental abundances were adopted, given
as number densities with respect to hydrogen: He/H = 0.1,
C/H = 1.4$\times$10$^{-4}$, N/H = 8.32 $\times$ 10$^{-5}$, O/H = 3.2 $\times$ 10$^{-4}$,
Ne/H = 1.2 $\times$ 10$^{-4}$, Mg/H = 1.1 $\times$ 10$^{-6}$, Si/H = 1.7 $\times$ 10$^{-6}$,
S/H = 2.8 $\times$ 10$^{-5}$. These are solar abundances (Asplund et al.
2005) depleted according to Savage \& Sembach (1996). 
D'Alessio et al. (2001) consider discs composed of a bimodal dust
distribution, where atmospheric dust follows the standard MRN model
and interior dust consists of larger grains with a size distribution
still described by a power law of index -3.5, but with a$_{min}$ = 0.005
$\mu$m and a$_{max}$ = 1 mm. The transition between atmospheric and interior
dust occurs at a height of 0.1 times the midplane gas scale height. As
we are mainly interested in regions well above the transition point, we
have chosen for simplicity to use atmospheric dust everywhere.
Following D'Alessio et al. (2001), the dust-to-gas mass ratio of
graphite is 0.00025 and that of silicates is 0.0004.

We refer the reader to Ercolano et al. (2008b, 2009a) and O10 for
further details about the codes used and the model setup.

\subsection{Differences in photoevaporation models}

All photoevaporation models are based on the idea that a source of
radiation (internal - e.g. the YSO, or external e.g. a nearby OB star)
heats up the gas in the disc, depositing enough energy to leave the
gas unbound. Low-mass YSOs that are far from external sources of
energetic radiation, e.g. a typical YSO in Taurus, are the only source
of irradiation for their discs. A typical YSO produces chromospheric
EUV (13.6~eV $\le$ E $<$ 100~eV) and X-ray radiation (E $>$ 100~eV),
with luminosities for the latter spanning several orders of magnitude
around $\sim10^{30}$erg/sec. The EUV luminosities are poorly known as
they cannot be directly observed, but they are estimated to be of order
10$^{41}$ phot/sec (Alexander et al. 2005). The YSO also radiates copiously in the FUV, both %reference Richard's paper
from the chromosphere and due to accretion shocks (which may also
contribute some of the X-ray emission). Chromospheric FUV
luminosities are comparable to the observed X-ray luminosities and
accretion luminosities proportional to the accretion rate of the
object. 

Theoretical photoevaporation models that account for one or a combination
of EUV, X-ray and FUV have been constructed by a number of authors and
it is worthwhile to 
point out some of the basic differences amongst the current models. 
In terms of basic physics, the models can be split into two %changed simple to basic
categories: (i) Diffuse field and (ii) direct field dominated. %capatilized Diffuse

The standard primordial disc EUV models of Hollenbach et al 
(1994), Clarke et al. (2001) and Font et al. (2004) belong to the first
category. In these models the heating of the upper layers of the gaseous disc
is dominated by recombination photons from within the flow
itself. The mass loss rates in this case are accurately determined by
hydrodynamical calculations and
more importantly are the mass-loss rate profiles (Font et
al. 2004). Mass-loss rates are found to be of order 10$^{-10}$
M$_{\odot}$/yr with 
only a square root 
dependence on the ionising luminosity. The mass-loss radial profile is   
narrowly peaked at a few AU for solar type stars.

As a result of the combined effect of EUV photoevaporation and viscous
draining, most discs would eventually develop an inner hole when the
accretion rates fall to a level comparable to the wind rates (Clarke et al. 2001, Alexander et al. 2006a). % removed as in the models of and added cathie's paper.
At this point the models
become direct field dominated (i.e. dominated by radiation from the
central source) as the inner disc is removed and stellar EUV %added euv
photons can directly reach the inner edge of the photoevaporating
disc. 

The FUV photoevaporation model of Gorti \& Hollenbach (2009) and the
X-ray photoevaporation models of Ercolano et al (2008b,2009a) and Owen
et al (2010a) also fall in the direct field dominated category, both
during their primordial and inner hole phase. For models in this
category, the heating is more or less set by the direct flux from the
central source.%added brackets about chemistry 

Determination of accurate mass-loss rates relies on the ability to %re-written this sentance slightly
solve the radiative transfer and thermal calculation problem
within a hydrodynamical calculation. This has only been done so far for
the EUV model (Clarke et al 2001, Font et al 2004, Alexander et al
2006) and the X-ray model (Owen et al 2010), both in the primordial
and inner hole phase. A hydrodynamical calculation has not been
performed for the FUV model and mass-loss rates in this case have been
estimated 
by way of an approximation (Gorti \& Hollenbach 2009), meaning that the
values obtained carry potentially large uncertainties. 

\begin{figure*}
\begin{center}
\includegraphics[width=18cm]{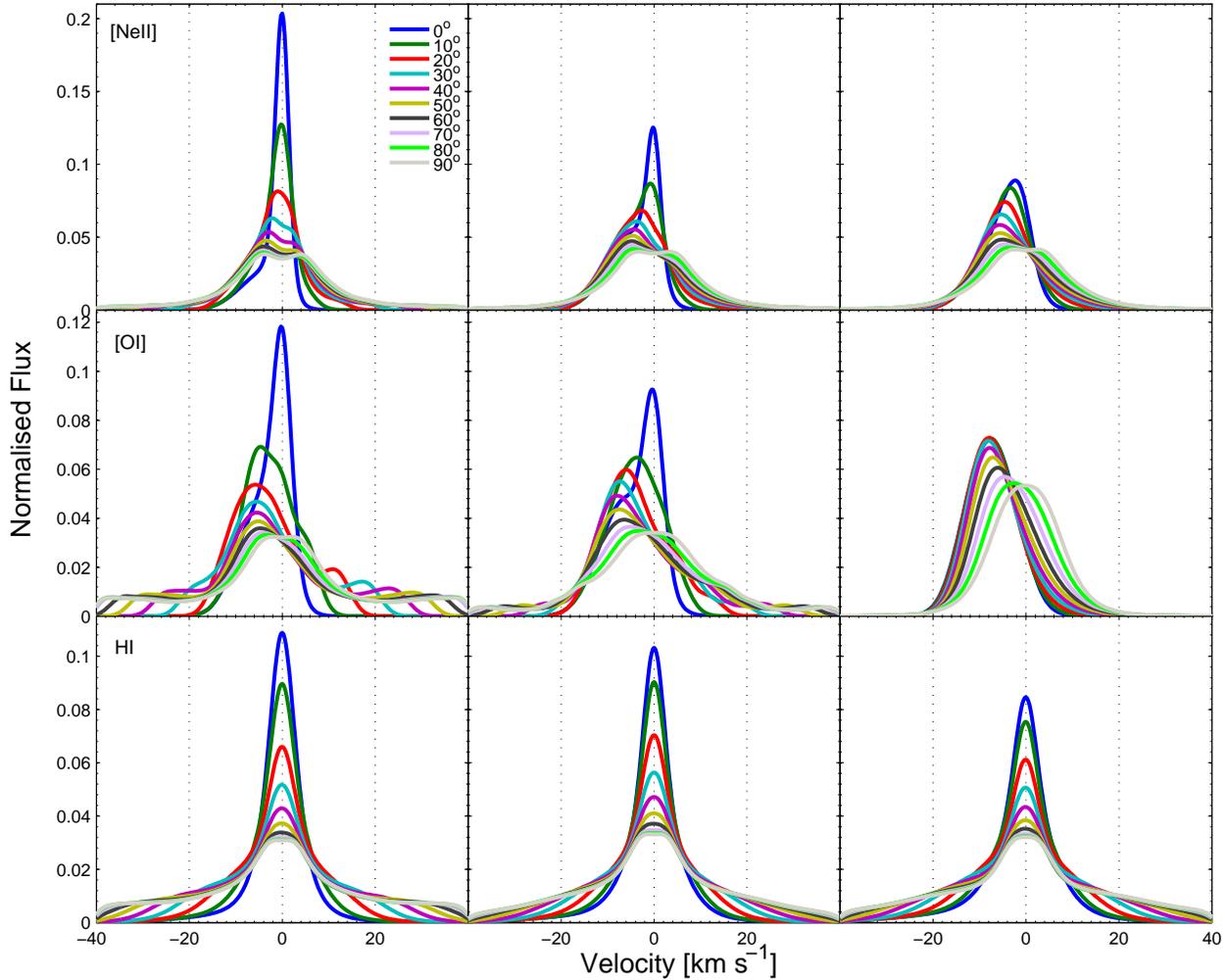}
\caption[]{Line profiles for (from the top) NeII~12.8$\mu$m,
  OI~6300{\AA} and a generic hydrogen recombination
  line. The left, middle and right panels are for primordial disc
  models irradiated by Log(L$_{X}$) = 28.3, 29.3 and 30.3,
  respectively. These line profiles are all normalised, such that
  $\int_{-40}^{40} L[v/(\textrm{km s}^{-1})]~\textrm{d}v=1$. The
  profiles are colour-coded according to their inclination angle, a
  colour legend in given in the top-left panel.}
\label{f1}
\end{center}
\end{figure*}

\subsubsection{Temperature determination}

The temperature structure in the upper layers of the gaseous disc is
crucial in determining the flow launching zone and mass loss rates
(hence the wind structure). The thermal balance of EUV and X-ray
irradiated models is set by heating by photoionisation against cooling
which is dominated by collisionally excited line emission. The physics
and atomic data involved are reasonably well known. Ercolano et al (2008b)
describes all the heating and cooling mechanisms for the X-ray model
in some detail.  
It is worth noting here that Ercolano et al (2008b, 2009a) and Owen et
al (2010a) use the {\sc  mocassin} code (Ercolano et al 2003, 2005,
2008a) which does not include molecular cooling in its thermal
balance. The lack of molecular cooling has been
criticised by Gorti \& Hollenbach (2009) as a possible source of
uncertainty.  However, we note that the launching region and the
associated sonic surface occurs in a %added sonic surface   
region where all molecules have already dissociated. While the X-rays
can deposit energy in the deep molecular layers, their largest thermal
impact is by far in the upper photodissociated layers. Furthermore,
even at large column densities, where molecular cooling can occur,
Figure~4 of Glassgold et al (2004) clearly shows that it is quickly
overwhelmed by dust-gas collisions, which are fully implemented in
{\sc mocassin}. We therefore conclude that the omission of molecular
cooling is not a significant source of uncertainty in our models (see
also further tests and discussion in Owen et al. 2010a).

In the FUV case explored by Gorti \& Hollenbach (2009) the heating of
the gas is dominated by photoelectric emission from dust grains, with
PAHs completely 
dominating the heating channels (see e.g. Figure 4 of Gorti \&
Hollenbach 2008). There are two problems with this approach. The
first is that the photoelectric yields are very poorly known, with 
uncertainties that could be as large as two orders of magnitude
(e.g. compare Abbas et al 2006, with Feuerbacher \& Fitton
1972). Another complication is the fact that the abundance (or even
presence) of PAH in T-Tauri discs is highly uncertain with PAHs
detections only in approximately 8\% of cases (e.g. Geers et al
2006). Indeed models 
predict that PAHs in the discs upper layers are easily destroyed
by X-ray irradiation (e.g. Siebenmorgen et al 2010). Even assuming
that photoelectric yields are correct, the uncertainty in the 
abundance/presence of PAHs in the disc makes photoelectric heating
almost a free parameter in these calculations and the temperature
structure largely uncertain.

\section{Results}

The results of our line luminosities and profile calculations (for a
subset of the lines) are presented in this section. We note that for
simplicity we choose not to follow the convention of using square brackets
around forbidden transitions. 

\begin{figure*}
\begin{center}
\includegraphics[width=18.cm]{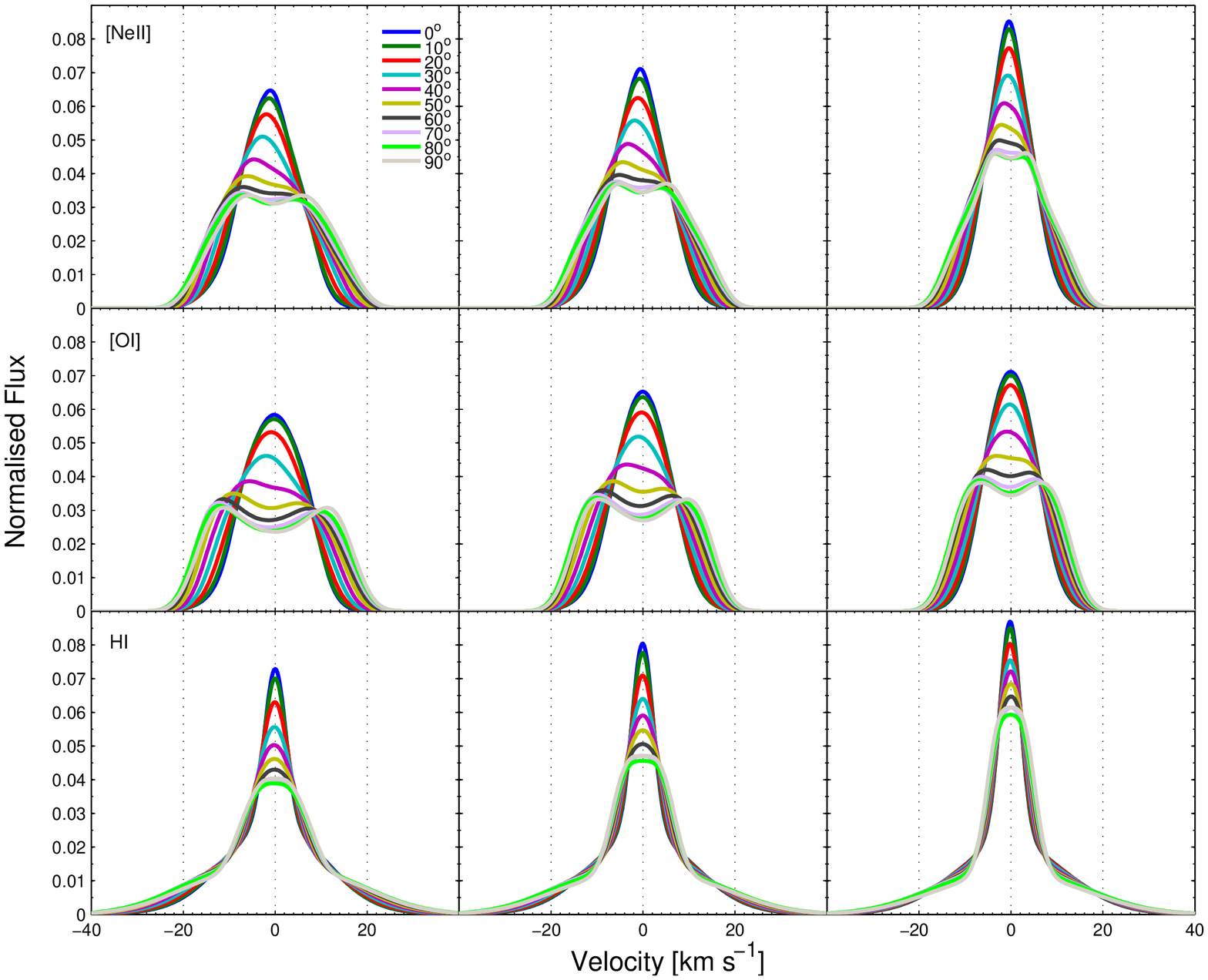}
\caption[]{Line profiles for (from the top) NeII~12.8$\mu$m,
   OI~6300{\AA} and a generic hydrogen recombination
  line. The left, middle and right panels are for inner-hole disc
  models irradiated by Log(L$_{X}$) = 30.3, with inner holes of radius
  8.3, 14.2 and 30.5,  respectively. These line profiles are all
  normalised, such that $\int_{-40}^{40} L[v/(\textrm{km
      s}^{-1})]~\textrm{d}v=1$. The
  profiles are colour-coded according to their inclination angle, a
  colour legend in given in the top-left panel.}
\label{f2}
\end{center}
\end{figure*}

\begin{figure*}
\begin{center}
\includegraphics[width=18.cm]{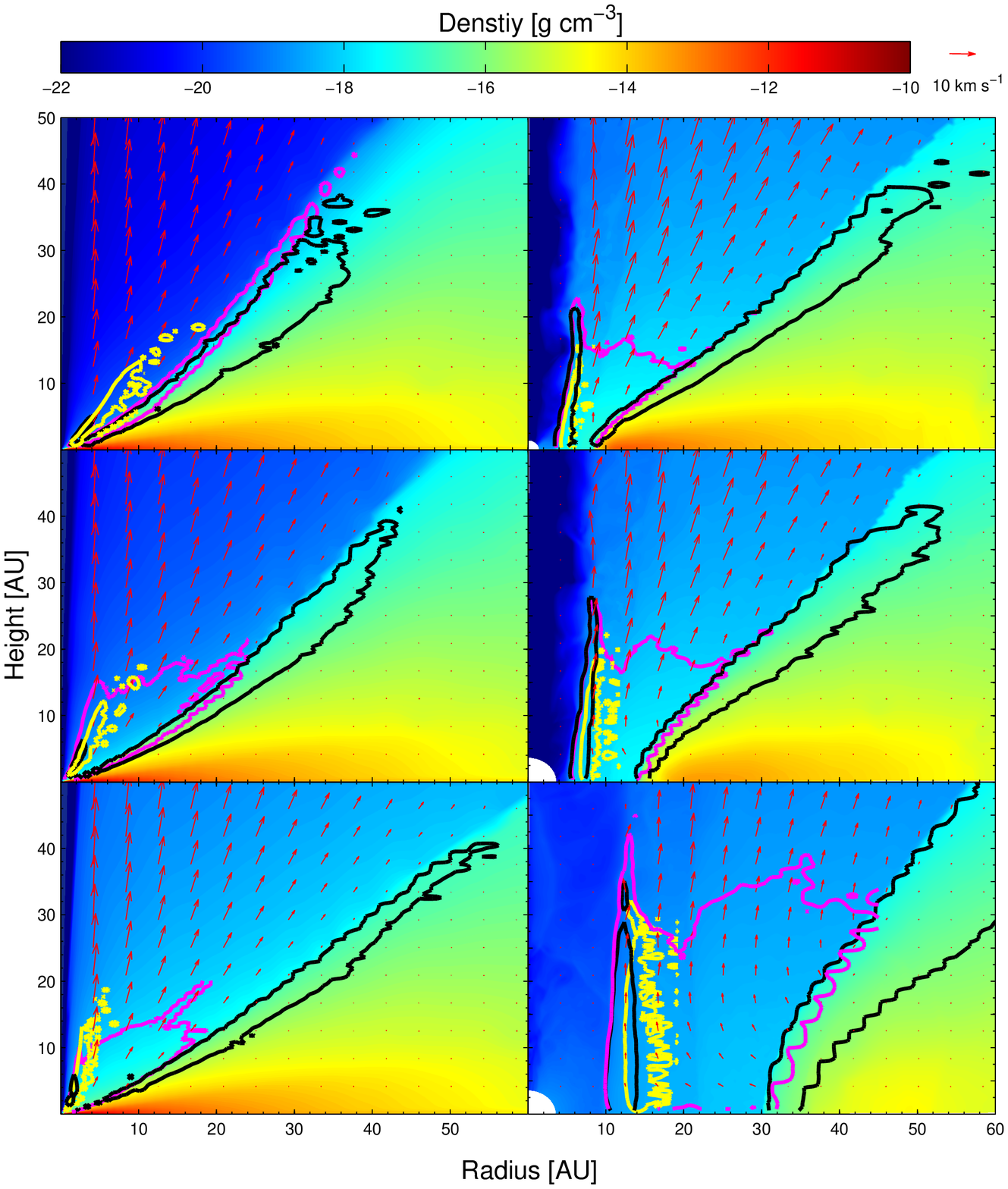}
\caption[]{Density maps showing the location of the 85\% emission region of HI recombination lines (black contour), OI~6300{\AA} line (yellow contour) and NeII~12.8~$\mu$m line (magenta contour). The velocity field is represented by the red arrows. {\it Left:} Primordial disc models irradiated by Log(L$_{X}$)~=~28.3~erg/sec (top), Log(L$_{X}$)~=~29.3~erg/sec (middle) and Log(L$_{X}$)~=~30.3~erg/sec (bottom). {\it Right:} Inner hole models irradiated by Log(L$_{X}$)~=~30.3~erg/sec with inner hole radii of R$_{in}$~=~8.3~AU,  R$_{in}$~=~14.2~AU and  R$_{in}$~=~30.5~AU}
\label{f3}
\end{center}
\end{figure*}
 
\subsection{Atlas of line intensities}
\label{s:a}
Tables~\ref{t1} to~\ref{t4} list the emission line spectra for the
models described in the previous section. The spectra were calculated
using the {\sc mocassin} code and using, as the input density structure, the %added the and comma
output of our modified {\sc zeus-2d} radiation-hydrodynamic
calculations. The disk atmosphere is optically thin in the listed
lines and therefore radiative transfer calculations in the lines is
not necessary. Only a selected
subset of lines is included here, however the full set is available upon
request from the authors. 

The tables are organised as follows. Tables~\ref{t1} and~\ref{t3} list
the luminosities of the collisionally excited transition of heavy 
elements and Tables~\ref{t2} and~\ref{t4} list a selection of hydrogen
and helium recombination
lines. Tables~\ref{t1} and~\ref{t2} show the results for the
primordial disc models with varying X-ray luminosity
(Log(L$_{X}$) = 28.3, 29.3 and 30.3 erg/sec) while
Tables~\ref{t3} and~\ref{t4} show the results for the inner hole
models irradiated by Log(L$_{X}$)~=~30.3~erg/sec and with varying
hole sizes.  

A casual inspection of these tables immediately
reveals that the integrated line luminosities are not very sensitive
(factors of two to five at most) to the
size of the inner hole for the range considered here
(8.3~AU~$\le$~R$_{in}~\le$~30.5~AU). However, inner hole
models do produce, in most cases, collisionally excited lines that are
approximately a factor two 
more luminous compared to those of a primordial disc with the same
X-ray luminosity. This is due to the different physical conditions
in inner hole winds that are less dense and most importantly on
average warmer than 
primordial disc winds.

Line luminosities are, in general, correlated with X-ray
luminosity, the response of different lines to changes in L$_{X}$
is however, different for different transitions, and depends on how
sensitive to changes in luminosity a line's emitting region is, and on
the dependance of a given emission
line to temperature, density or ionisation structure. 
 A number of factors are relevant to determine which region contributes to the
 emission of a given line. The ionic abundance is of course
 important; collisionally excited lines of NeII 
and OI require Ne$^+$ and O$^0$, as well as colliding particles, like
(e.g.) electrons and/or neutral hydrogen. HI recombination lines
arise in region where
there is a non zero abundance of H$^+$ and e$^-$. Another key property is
temperature. Collisionally excited lines depend exponentially on the gas
temperature through the Boltzmann term and therefore
preferentially trace the hotter regions. In the case of collisionally
excited lines in the infrared, however, this temperature dependence is
weakened by the fact that the gas temperature is generally much higher
than the line excitation temperature, causing the Boltzmann term to
tend to unity. Recombination lines on the other hand, only have an
inverse square root dependence on temperature and therefore are
emitted preferentially in cooler regions, as long as these are not
completely neutral. Gas density also influences the emissivity of collisionally
excited lines through critical density arguments.

It should also be noted that increasing the X-ray luminosity
not only affects the ionisation rate of the
wind, but also affects its density. Higher luminosities drive a more
vigorous (denser) wind, with a roughly linear correlation between
L$_X$ and gas density in the wind. For the primordial discs presented
here the scaling of the density 
with L$_X$ can be understood simply. Given the wind is thermally
driven, the specific energy requirements to drive the flow will
only depend on the local value of the effective potential. Once
a freely expanding wind is established it will remain at approximately
constant temperature. Since for transonic/sonic flows the
velocity is strongly coupled to the sound speed of the gas, which
in turn only depends on the value of the gas temperature, then, for all
density and temperature regimes we are interested in, this local
specific energy requirement is 
equivalent to a local temperature condition in the launch
region\footnote{We have and will \emph{implicitly} assume
that the total input energy from the high energy radiation is
sufficient to balance the work done driving any flow. See O10 of a
detailed discussion of this condition in relation to X-ray driven
photoevaporative winds.}. Thus the spatial variation of temperature
within the flow is approximately set by the
mass of the central star, and will be independent of the local
density and X-ray luminosity. 
Given also that temperature can be expressed as a monotonic function
of the ionisation parameter, the spatial variation of the
ionisation parameter is approximately fixed by the mass of the
central star. Hence, using the definition of the ionisation parameter
$\xi=$L$_X/nr^2$ and given the value of the ionisation parameter is
spatially fixed, at a point $r=$constant\footnote{Note throughout this paper we use $(r,\theta,\phi)$ and $(R,\varphi,z)$ to distinguish between spherical and cylindrical polar co-ordinates respectfully.} we find that gas density
$n\propto $L$_X$, as can be seen in the primordial discs shown in
Figure~\ref{f3}. %removed last paragraph as it's probably unecessay 

An element-by-element discussion, comments on observability of
the various transitions and a comparison with observations and other
theoretical investigation is postponed to Section~4

\begin{table*}
\caption[]{Collisionally excited lines. Primordial disk models with varying
X-ray luminosity. Wavelengths are in vacuum. Only selected
 lines are listed here, the full set is available upon request from the authors. }
\label{t1}
\begin{tabular}{cccccccccc}
\hline
Species & Wavelength  & \multicolumn{3}{c}{Line Luminosity  $[$L$_{\odot}]$}&Species & Wavelength  & \multicolumn{3}{c}{Line Luminosity  $[$L$_{\odot}]$}  \\
        &  $[$A$]$    & Log(L$_X$)=28.3  &29.3&30.3&&  $[$A$]$    & Log(L$_X$)=28.3  & 29.3&30.3 \\ 
\hline

CI  & 8729.51     &    1.79E-08 & 8.21E-08 & 5.78E-07  & MgI     &  4573.180   &6.11E-08 & 8.21E-07 & 1.08E-05 \\
CI  & 9826.85     &    4.32E-08 & 3.67E-08 & 6.23E-07  &   MgI     &  4564.670   &1.37E-09 & 4.73E-09 & 2.16E-08\\ 
CI  & 9852.99     &    1.27E-07 & 1.06E-08 & 1.16E-06  &   MgI     &  2852.140   &8.05E-09 & 1.16E-07 & 1.54E-06\\ 
CII &  1577287.   &    8.80E-08 & 2.80E-07 & 6.64E-07   &   MgII    &  2803.530   &4.77E-09 & 7.92E-08 & 1.28E-06\\ 
CII &  2325.400   &    3.13E-09 & 5.60E-08 & 9.43E-07   &   MgII    &  2796.350   &9.40E-09 & 1.54E-07 & 2.47E-06\\ 
CII  &  2324.210   &   3.13E-09 & 5.02E-08 & 5.87E-07  &    SiII   &  348189.4   & 5.67E-07 & 1.06E-06 & 1.85E-06\\ 
CII  &  1334.530   &   2.08E-10 & 4.35E-09 & 7.42E-08  &    SiII   &  2335.120   & 1.20E-10 & 2.03E-09 & 3.24E-08\\ 
CII  &  2328.840   &   3.73E-09 & 6.67E-08 & 1.11E-06  &    SiII   &  1808.010   & 1.43E-10 & 2.61E-09 & 4.17E-08\\ 
CII  &  2327.640   &   9.84E-09 & 1.54E-07 & 1.85E-06  &    SiII   &  2350.890   & 7.90E-11 & 1.35E-09 & 2.00E-08\\ 
CII  &  2326.110   &   2.08E-08 & 3.57E-07 & 5.56E-06  &    SiII   &  2344.920   & 3.87E-10 & 6.57E-09 & 1.00E-07\\ 
CII  &  1335.660   &   4.02E-11 & 8.60E-10 & 1.48E-08  &    SiII   &  2335.320   & 5.52E-10 & 9.57E-09 & 1.46E-07\\ 
CII  &  1335.710   &   3.58E-10 & 7.83E-09 & 1.34E-07  &    SiII   &  1816.930   & 2.53E-10 & 4.64E-09 & 7.26E-08\\ 
OI   &  6302.030   &   2.83E-07 & 1.93E-06 & 1.25E-05  &    SII    &  6732.690   & 2.08E-07 & 1.16E-06 & 1.08E-05\\ 
OI   &  6365.530   &   9.10E-08 & 6.28E-07 & 4.01E-06  &    SII    &  6718.310   & 1.41E-07 & 7.44E-07 & 5.72E-06\\ 
OI   &  2973.160   &   6.11E-10 & 9.18E-09 & 1.06E-07  &    SII    &  4077.510   & 5.81E-08 & 4.25E-07 & 3.24E-06\\ 
OI   &  5578.890   &   5.22E-09 & 7.83E-08 & 9.12E-07  &    SII    &  4069.760   & 2.23E-07 & 1.64E-06 & 1.22E-05\\ 
OII   &  3729.880   &  1.79E-08 & 1.06E-07 & 6.49E-07  &    SII    &  10339.23   & 3.87E-08 & 2.80E-07 & 2.16E-06\\ 
OII   &  3727.090   &  3.58E-08 & 2.51E-07 & 1.85E-06  &    SII    &  10289.55   & 5.52E-08 & 4.06E-07 & 2.93E-06\\ 
OII   &  2470.970   &  2.23E-09 & 2.70E-08 & 2.78E-07  &    SII    &  10373.34   & 1.64E-08 & 1.16E-07 & 9.12E-07\\ 
OII   &  2471.090   &  6.86E-09 & 8.70E-08 & 8.65E-07  &    SII    &  10323.33   & 6.56E-08 & 4.83E-07 & 3.55E-06\\ 
OII   &  7320.910   &  1.79E-09 & 2.12E-08 & 2.16E-07  &    SIII   &  335008.4   & 1.64E-08 & 1.25E-07 & 1.28E-06\\ 
OII   &  7321.980   &  4.17E-09 & 5.41E-08 & 5.41E-07  &    SIII   &  187055.7   & 2.38E-08 & 1.64E-07 & 2.00E-06\\ 
OII   &  7331.700   &  2.83E-09 & 3.57E-08 & 3.55E-07  &    SIII   &  9070.050   & 3.58E-09 & 5.41E-08 & 8.19E-07\\ 
OII   &  7332.780   &  2.23E-09 & 2.90E-08 & 2.93E-07  &    SIII   &  3722.450   & 2.23E-10 & 5.02E-09 & 8.34E-08\\ 
NeII   &  128155.8   & 9.25E-08 & 5.51E-07 & 5.41E-06  &    SIII   &  9532.250   & 2.08E-08 & 3.09E-07 & 4.63E-06\\ 
NeIII   &  155545.2   &1.01E-08 & 1.16E-07 & 1.06E-06  &    SIII   &  1728.950   & 2.23E-11 & 7.73E-10 & 2.16E-08\\ 
NeIII   &  3869.850   &2.98E-09 & 6.28E-08 & 8.34E-07  &    SIII   &  6313.650   & 4.02E-10 & 8.89E-09 & 1.48E-07\\ 
NeIII   &  360230.6   &5.96E-10 & 6.47E-09 & 5.72E-08  &    SIV    &  105108.3   & 4.92E-08 & 2.99E-07 & 3.71E-06\\ 
NeIII   &  3968.580   &8.80E-10 & 1.93E-08 & 2.47E-07  &           &             &          &          &         \\
\hline                                     
\end{tabular}                              
\end{table*}     

\begin{table}
\caption[]{Recombination Lines of H, HeI and HeII. Primordial disc models with varying X-ray
  luminosity. Wavelengths are in vacuum. 
Only selected lines are listed here, the full set is available upon request from the authors. }
\label{t2}
\begin{tabular}{ccccc}
\hline
\multicolumn{5}{c}{ Hydrogen }\\
Transition & Wavelength  & \multicolumn{3}{c}{Line Luminosity $[$L$_{\odot}]$}\\
&  $[$A$]$    & Log(L$_X$)=28.3 & 29.3  &30.3 \\ 
\hline
\multicolumn{5}{c}{ HI }\\
 3-2 &  6564.696  &    2.62E-07 & 2.77E-06 & 1.10E-05  \\
 4-2 &  4862.738  &    7.87E-08 & 9.18E-07 & 3.20E-06  \\
 5-2 &  4341.730  &    3.52E-08 & 4.26E-07 & 1.43E-06  \\
 6-2 &  4102.935  &    1.94E-08 & 2.41E-07 & 7.97E-07  \\
 5-3 &  12821.67  &    1.62E-08 & 1.67E-07 & 6.70E-07  \\
 6-3 &  10941.16  &    8.38E-09 & 9.13E-08 & 3.40E-07  \\
 7-3 &  10052.19  &    5.03E-09 & 5.70E-08 & 2.04E-07  \\
 8-3 &  9548.649  &    3.35E-09 & 3.91E-08 & 1.35E-07  \\
 7-4 &  21661.29  &    3.05E-09 & 3.01E-08 & 1.26E-07  \\
 7-5 &  46537.93  &    2.20E-09 & 1.88E-08 & 9.64E-08  \\
 7-6 &  123719.3  &    1.68E-09 & 1.23E-08 & 7.99E-08  \\
 8-7 &  190619.3  &    7.79E-10 & 5.50E-09 & 3.76E-08  \\
 9-7 &  113087.2  &    5.07E-10 & 3.86E-09 & 2.35E-08  \\
\multicolumn{5}{c}{ HeI }\\
4$^3$D-2$^3$P &  4473.000   &   8.17E-10 & 1.63E-08 & 3.50E-08  \\
3$^3$P-2$^3$S &  3890.000   &   1.81E-09 & 3.79E-08 & 8.50E-08  \\
3$^3$D-2$^3$P &  5877.000   &   2.38E-09 & 4.79E-08 & 1.03E-07  \\
2$^3$P-2$^3$S &  10833.00   &   1.82E-08 & 4.43E-07 & 1.10E-06  \\
\multicolumn{5}{c}{ HeII }\\
 4-3 &  4689.069  & 1.77e-10 & 3.91e-09 & 5.92e-08 \\
\hline

\end{tabular}
\end{table}

\begin{table*}
\caption[]{Collisionally excited lines. Inner hole disk models with varying
hole size. Wavelengths are in vacuum. Only selected
 lines are listed here, the full set is
 available upon request from the authors. }
\label{t3}
\begin{tabular}{cccccccccc}
\hline
Species & Wavelength  & \multicolumn{3}{c}{Line Luminosity $[$L$_{\odot}]$}&Species & Wavelength  & \multicolumn{3}{c}{Line Luminosity $[$L$_{\odot}]$} \\
        &  $[$A$]$    & R$_{in}$=8.3~AU  &14.2~AU & 30.5~AU & & $[$A$]$    & R$_{in}$=8.3~AU  &14.2~AU & 30.5~AU  \\ 
\hline
CI          & 8729.51   &    5.57E-07&  5.57E-07&  5.38E-07& NeIII      & 3869.850  &    2.25E-06&  1.99E-06&  2.61E-06\\
CI          & 9826.85   &    7.52E-07&  5.57E-07&  7.02E-07& NeIII      & 360230.6  &    1.21E-07&  1.57E-07&  2.24E-07\\
CI          & 9852.99   &    1.30E-06&  1.66E-06&  2.09E-06& NeIII       & 3968.580  &    6.86E-07&  6.14E-07&  7.48E-07\\
CII         & 1577287.  &    1.12E-06&  1.41E-06&  2.09E-06& NeIV        & 2422.360  &    1.56E-08&  1.90E-09&  8.98E-08\\
CII         & 2325.400  &    5.03E-07&  4.48E-07&  3.96E-07& MgI         & 4573.180  &    1.12E-06&  9.13E-07&  6.66E-07\\
CII         & 2324.210  &    5.38E-07&  4.81E-07&  4.26E-07& MgI         & 4564.670  &    5.99E-08&  7.22E-08&  9.72E-08\\
CII         & 1334.530  &    1.04E-07&  9.13E-08&  8.98E-08& MgI         & 2852.140  &    2.69E-07&  2.32E-07&  1.87E-07\\
CII         & 2328.840  &    5.99E-07&  5.31E-07&  4.71E-07& MgII        & 2803.530  &    1.04E-06&  9.13E-07&  7.48E-07\\
CII         & 2327.640  &    1.65E-06&  1.49E-06&  1.27E-06& MgII        & 2796.350  &    1.99E-06&  1.82E-06&  1.49E-06\\
CII         & 2326.110  &    3.30E-06&  2.90E-06&  2.61E-06& SiII        & 348189.4  &    1.65E-06&  1.66E-06&  1.87E-06\\
CII         & 1335.660  &    1.99E-08&  1.74E-08&  1.79E-08& SiII        & 2335.120  &    1.99E-08&  1.82E-08&  1.64E-08\\
CII         & 1335.710  &    1.82E-07&  1.57E-07&  1.64E-07& SiII        & 1808.010  &    4.08E-08&  3.65E-08&  3.59E-08\\
OI          & 6302.030  &    4.51E-05&  4.48E-05&  3.59E-05& SiII        & 2350.890  &    1.30E-08&  1.16E-08&  1.12E-08\\
OI          & 6365.530  &    1.39E-05&  1.41E-05&  1.12E-05& SiII        & 2344.920  &    6.42E-08&  5.72E-08&  5.31E-08\\
OI          & 2973.160  &    1.73E-08&  1.07E-08&  5.98E-09& SiII        & 2335.320  &    9.55E-08&  8.30E-08&  7.48E-08\\
OI          & 5578.890  &    1.47E-07&  9.13E-08&  5.16E-08& SiII        & 1816.930  &    7.21E-08&  6.39E-08&  6.28E-08\\
OII         & 3729.880  &    1.91E-06&  2.32E-06&  3.44E-06& SII         & 6732.690  &    2.78E-05&  3.07E-05&  3.14E-05\\
OII         & 3727.090  &    5.47E-06&  6.64E-06&  8.98E-06& SII         & 6718.310  &    1.47E-05&  1.57E-05&  1.87E-05\\
OII         & 2470.970  &    9.55E-07&  9.13E-07&  7.40E-07& SII         & 4077.510  &    7.73E-06&  6.64E-06&  5.01E-06\\
OII         & 2471.090  &    3.30E-06&  3.15E-06&  2.76E-06& SII         & 4069.760  &    2.43E-05&  1.99E-05&  1.49E-05\\
OII         & 7320.910  &    7.73E-07&  6.97E-07&  5.91E-07& SII         & 10339.23  &    5.12E-06&  4.40E-06&  3.29E-06\\
OII         & 7321.980  &    2.08E-06&  1.99E-06&  1.72E-06& SII         & 10289.55  &    5.90E-06&  4.89E-06&  3.59E-06\\
OII         & 7331.700  &    1.30E-06&  1.16E-06&  9.72E-07& SII         & 10373.34  &    2.08E-06&  1.82E-06&  1.34E-06\\
OII         & 7332.780  &    1.12E-06&  1.07E-06&  8.98E-07& SII         & 10323.33  &    6.86E-06&  5.72E-06&  4.19E-06\\
OIII        & 883392.2  &    7.03E-08&  7.72E-08&  1.12E-07& SIII        & 335008.4  &    1.82E-06&  1.99E-06&  2.32E-06\\
OIII        & 518134.7  &    2.78E-07&  3.23E-07&  4.71E-07& SIII        & 8830.960  &    1.65E-08&  1.41E-08&  1.49E-08\\
OIII        & 4960.290  &    5.38E-07&  3.98E-07&  1.87E-06& SIII        & 187055.7  &    3.38E-06&  3.65E-06&  4.04E-06\\
OIII        & 2321.670  &    1.39E-08&  6.22E-09&  6.88E-08& SIII        & 9070.050  &    1.91E-06&  1.66E-06&  1.79E-06\\
OIII        & 5008.240  &    1.65E-06&  1.16E-06&  5.53E-06& SIII        & 3722.450  &    1.30E-07&  1.07E-07&  1.49E-07\\
OIII        & 1666.150  &    1.91E-08&  7.30E-09&  1.12E-07& SIII        & 9532.250  &    1.12E-05&  9.13E-06&  1.04E-05\\
OIII        & 4364.450  &    1.56E-08&  7.05E-09&  7.48E-08& SIII        & 1728.950  &    2.51E-08&  1.99E-08&  4.86E-08\\
NeII        & 128155.8  &    1.21E-05&  1.16E-05&  1.04E-05& SIII        & 6313.650  &    2.34E-07&  1.82E-07&  2.61E-07\\
NeIII       & 155545.2  &    2.43E-06&  2.90E-06&  3.44E-06& SIV         & 105108.3  &    5.73E-06&  5.64E-06&  5.08E-06\\
\hline                                     
\end{tabular}                              
\end{table*}     

\begin{table}
\caption[]{Recombination lines of H, HeI and HeII. Inner hole disc models with
  varying hole size. Wavelengths are in vacuum. 
Only selected lines are listed here, the full set is
 available upon request from the authors. }
\label{t4}
\begin{tabular}{ccccc}
\hline
Transition & Wavelength  & \multicolumn{3}{c}{Line Luminosity $[$L$_{\odot}]$} \\
           &  $[$A$]$    & 8.3  &  14.2    &  30.5  \\ 
\hline
  3-2&   6564.696&      1.40E-05&  2.43E-05&  4.76E-06 \\  
  4-2&   4862.738&      4.58E-06&  7.88E-06&  1.55E-06 \\  
  5-2&   4341.730&      2.10E-06&  3.60E-06&  7.11E-07 \\  
  6-2&   4102.935&      1.15E-06&  1.97E-06&  3.90E-07 \\  
  5-3&   12821.67&      8.28E-07&  1.44E-06&  2.81E-07 \\  
  6-3&   10941.16&      4.42E-07&  7.67E-07&  1.49E-07 \\  
  7-3&   10052.19&      2.67E-07&  4.61E-07&  9.03E-08 \\  
  8-3&   9548.649&      1.75E-07&  3.02E-07&  5.90E-08 \\  
  7-4&   21661.29&      1.47E-07&  2.59E-07&  5.02E-08 \\  
  7-5&   46537.93&      9.62E-08&  1.71E-07&  3.31E-08 \\  
  7-6&   123719.3&      6.62E-08&  1.20E-07&  2.30E-08 \\  
  8-7&   190619.3&      2.98E-08&  5.42E-08&  1.03E-08 \\  
  9-7&   113087.2&      2.04E-08&  3.70E-08&  7.12E-09 \\  
\multicolumn{5}{c}{ HeI }\\
4$^3$D-2$^3$P &  4473.000  & 2.45E-07 & 2.42E-07 & 2.44E-07 \\
3$^3$P-2$^3$S &  3890.000  & 6.09E-07 & 5.94E-07 & 5.98E-07 \\
3$^3$D-2$^3$P &  5877.000  & 7.21E-07 & 7.11E-07 & 7.18E-07 \\
2$^3$P-2$^3$S &  10833.00  & 7.54E-06 & 6.99E-06 & 6.72E-06 \\
\multicolumn{5}{c}{ HeII }\\
 4-3 &  4689.069  & 1.21031e-07 &  1.08842e-07 & 2.01123e-07\\
\hline      
\end{tabular}
\end{table}

\subsection{Line profiles}

\begin{figure*}
\begin{center}
\includegraphics[width=18.cm]{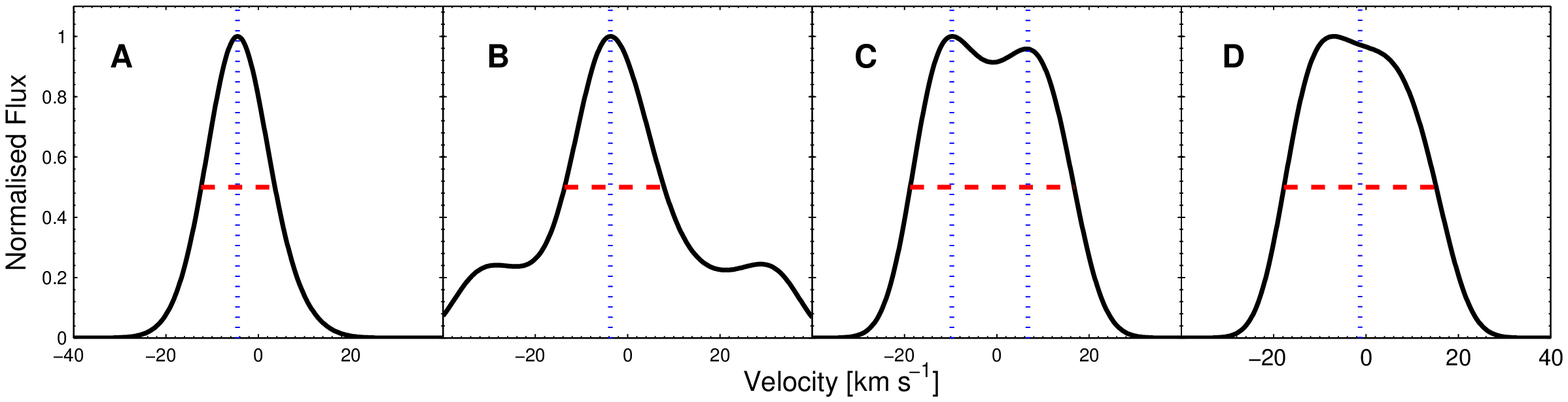}
\caption[]{Representative shapes for the line profile shapes
  calculated from our models. Profile A (left) can be well
  approximated by a Gaussian. Profile B is a Gaussian with large %replaced comma with full stop
  wings. Profile C is a double peaked profile and profile D are profiles where %ditto
  two clear peaks are not resolved, yet the profile is broad,
  flat-topped or significantly asymmetric. The blue vertical dots 
  show the location of v$_{peak}$ as given in Table~\ref{t5} and the
  red horizontal dashed line shows the FWHM for the given profiles.} 
\label{f4}
\end{center}
\end{figure*}

We have used our two-dimensional map of emissivities and gas
velocities to reconstruct a three-dimensional cube of the disc and
calculate the line-of-sight emission profiles for a number of
transitions. The emissivities are calculated using 120 radial points
($N_R=120$) and 1500 height points ($N_z=1500$). By assuming
azimuthal symmetry about the $Z$-axis, and reflection symmetry
about the disc mid-plane the resultant 3D grid has dimensions $N_R
\times 2N_z \times N_\phi$. We adopted a value of $N_\varphi=130$ and
checked that this was large enough to fully resolve the profiles. 
For the chosen emission lines the disc atmosphere is optically thin and
therefore we assume that the line contribution of each parcel of gas
can escape freely along the line of sight, provided that this does not
intercept the midplane (e.g. Alexander 2008, Schisano et
al. 2010). For almost edge-on inclinations escape columns are large
enough for dust attenuation to start playing a role. This is not
accounted for in this work and the magnitude of this effect depends
largely on poorly known parameters such as dust to gas ratios, the
degree of dust settling and grain growth in the disc.  

The line luminosity is then computed by including a Doppler broadening
term in each cell. Hence the line luminosity at a given velocity $v$
is computed by evaluating the following integral by direct summation. %changed can be to is to keep correct present tense
\begin{equation}
  L(v)=\int\limits_V\textrm{d}^3r \frac{\ell(r)}{\sqrt{2\pi
      v_{th}(r)^2}}\exp\left(-\frac{[v-v_{los}(
        r)]^2}{2v_{th}(r)^2}\right)
\end{equation}  
Where $\ell(r)$ is the volume averaged power emitted at a point
$r$, $v_{los}(r)$ is the projected gas velocity along the
line of sight and $v_{th}$ is the local thermal velocity (rms velocity) of the
emitting atom given by
\begin{equation}
  v_{th}=\sqrt{\frac{3K_BT(r)}{\mu_im_H}}
\end{equation}
With $\mu_i$ the mean molecular weight of the required atom.  
The resolution of the velocity array is 0.25~km/s.

We have evaluated the profiles of the OI~6300{\AA}, the
NeII~12.8$\mu$m and a generic HI recombination
line, as these have been the focus of recent observational studies
(e.g. HEG95, Pascucci \& Sterzik 2009; Najita et al 2009). Profiles
for any of the lines listed in the table (and 
many others that are not) can easily be
calculated upon request by the authors. 

Figure~\ref{f1} shows the normalised profiles of the
NeII fine structure line at 12.8$\mu$m, the OI forbidden line at
6300~{\AA} and a hydrogen 
recombination line for the primordial disc model
with Log(L$_{X}$)~=~28.3~erg/s (left), 29.3~erg/s (middle) and
30.3~erg/s (right). Figure~\ref{f2} shows the same line profiles for
inner-hole models with Log(L$_{X}$)~=~30.3~erg/s and inner hole
radii of 8.3~AU (left), 14.2~AU (middle) and 30.5~AU (right). Profiles
are computed for ten inclination angles from 0$^o$ to 
90$^o$ degrees in steps of 10$^o$. The profiles were degraded to an
instrumental resolution  of R~=~100000, typical of current ground based
facilities. 
%{\bf We note however that the lines are close to being fully  
%resolved at R~=~30000 and therefore the full-width-half-maximum (FWHM)
%would be  approximately the same at higher (e.g. 100000) resolution.}

Figure~\ref{f3} shows 2D density maps (colour gradients) of the primordial
disc (left panels) and inner-hole models (right panels), respectively,
with velocity vectors overlain. The models from top to bottom are for
increasing X-ray luminosities (left) and increasing inner hole
radii (right). The black, magenta and yellow contours delineate the region
containing 85\% of the emission for the HI recombination lines, the
NeII~12.8$\mu$m\footnote{The NeIII~15.5$\mu$m region is very similar.}
and the OI~6300{\AA}, respectively. The noisy appearance of some of
the emitting regions are due to Monte Carlo fluctuations and the
imposition of a hard cut-off to draw the contours. We have checked
that the noise does not affect luminosities and line profiles to more
than a 5\% level.

The OI~6300{\AA} line (yellow contours) originates almost exclusively
in the wind or in the very inner (hot and bound-$R<1$~AU) region
of the disc 
(in the primordial case), which is where the temperatures are high
enough to populate its upper level. The same would be true for all
collisionally excited optical and UV lines. It is the contribution
from the bound hot inner disc that produces the 'winged' profiles for
the OI~6300{\AA} clearly seen in the lower luminosity 
primordial models. 

The NeII mid-infrared (MIR) lines can arise
in much cooler regions and therefore, while still predominantly
produced in the wind, they are not confined to the hot inner layers,
but span over a larger region. The NeII emitting region
(magenta contours in Figure~\ref{f3}) and line profiles vary
significantly with X-ray luminosity, tracing the %replace baciscally following with tracing
highest electron density regions (L(NeII)~$\propto$n$_e^2$). 

The profiles of HI recombination lines are more complicated as they
have contributions arising both from the wind and from the bound
disc. However, as will be discussed later, our models underpredict the
luminosities of observed IR hydrogen recombination lines by one or two
orders of magnitude (Pascucci et al 2007; Ratzka et al 2007; Najita et
al 2010) and underpredict the luminosities of observed
  Pa$\beta$ and Br$\gamma$ lines by four orders of magnitude
  (Muzerolle et al, 1998), implying that the 
emission must be 
dominated by regions that are not included in our models (e.g. dense
plasma close to the star). The resulting line profiles may therefore
bear little resemblance to those shown in this paper. 
The left panels of Figure~\ref{f3} (black contours) show that in the
primordial disc models there is only a small wind contribution, which
results in the hydrogen recombination profiles showing no blue-shift at%replace H with hydrogen
all. While inner hole 
models have larger wind contributions to the HI recombination lines,
the resulting profiles are still not blue-shifted, although this is
virtually true for all line profiles from inner hole models, even
those formed exclusively in the wind. %Removed explanation mark.

The reason why inner hole models produce very small or no blue-shifts
is that at low inclinations both the blue and red-shifted components
of the wind are visible through the hole, resulting in only a small,
often undetectable, blue-shift. At higher inclination the lines are
Keplerian and no blue-shift is expected anyway. We will discuss the
significance of this result in the next section, when comparisons are made with %re-written this sentance
observations, but we note at this point that this is only true if the
winds are optically thin to the lines in question. 

To aid comparison with observations we have summarised the main
profile characteristics in Tables~\ref{t5} and~\ref{t6}. These include
the velocity of the peak(s), $v_{peak}$, the full-width-half-maxima
(FWHM) and the 
profile ``type''. We have classified the profiles with four main
types as shown in Figure~\ref{f4}. Type ``A'' is a single peaked
profile that could be well approximated by a Gaussian. Type ``B'' is 
a single peaked Gaussian-like profile, but with large %removed still
wings. This is generally produced when there is both a wind and bound
disc contribution to the line and/or the emission region is spread
over a large radial domain. Type ``C'' is a double peaked
profile, in this case the two values of $v_{peak}$ are given. Type
``D'' are profiles where two clear peaks are not resolved, yet the
profile is broad, flat-topped or significantly asymmetric; in this
case the value at the centre of the line (i.e. the middle of the FWHM
segment, horizontal dashed red line in Figure~\ref{f4}) is given;
strictly speaking,  this
is not the location of the 'peak' however we keep the label $v_{peak}$
for simplicity.  

\subsection{NeII~12.8$\mu$m and OI~6300{\AA} radial profiles}

\begin{figure*}
\begin{center}
\begin{minipage}[]{18cm}
\includegraphics[width=9.cm]{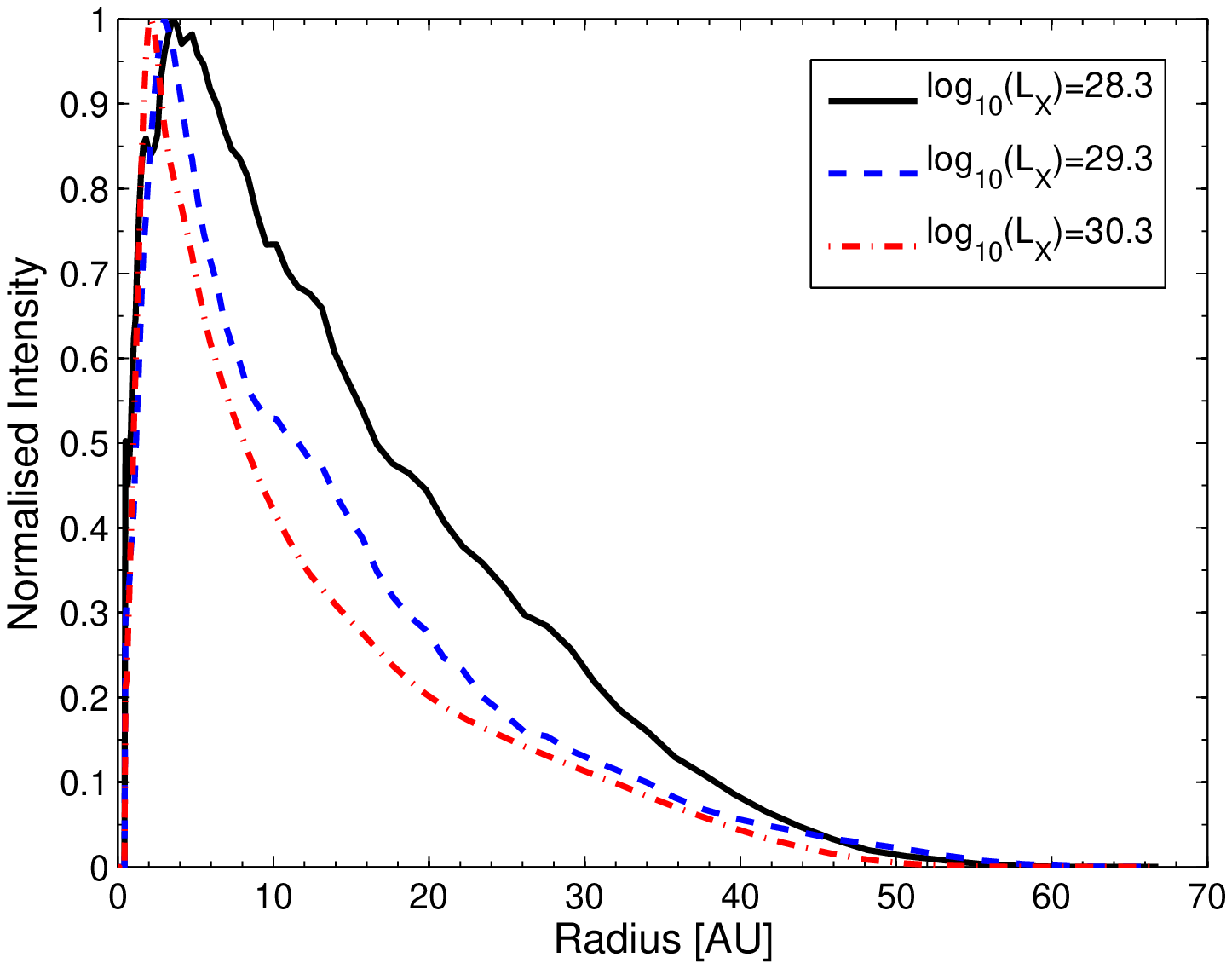}
\includegraphics[width=9.cm]{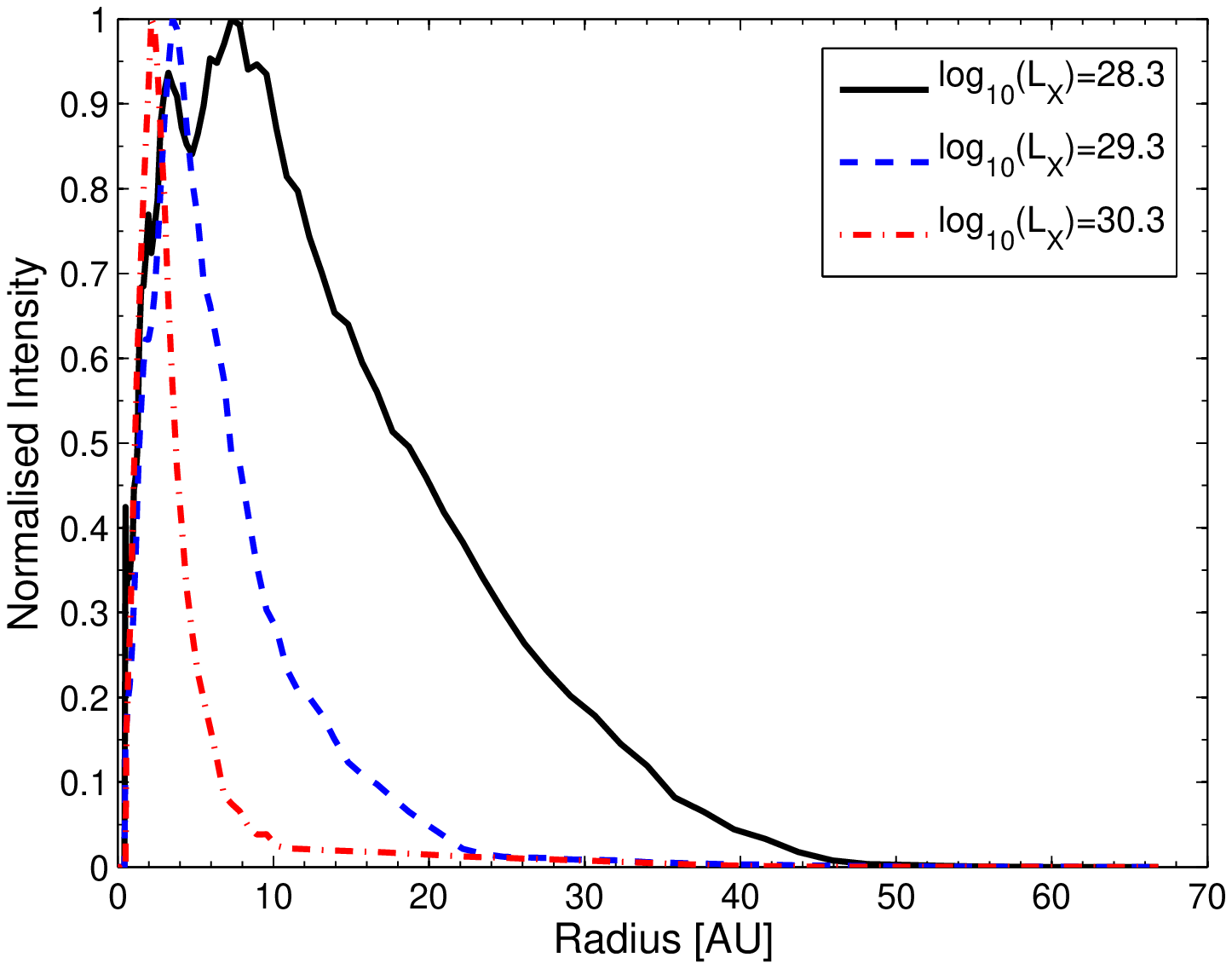}
\includegraphics[width=9.cm]{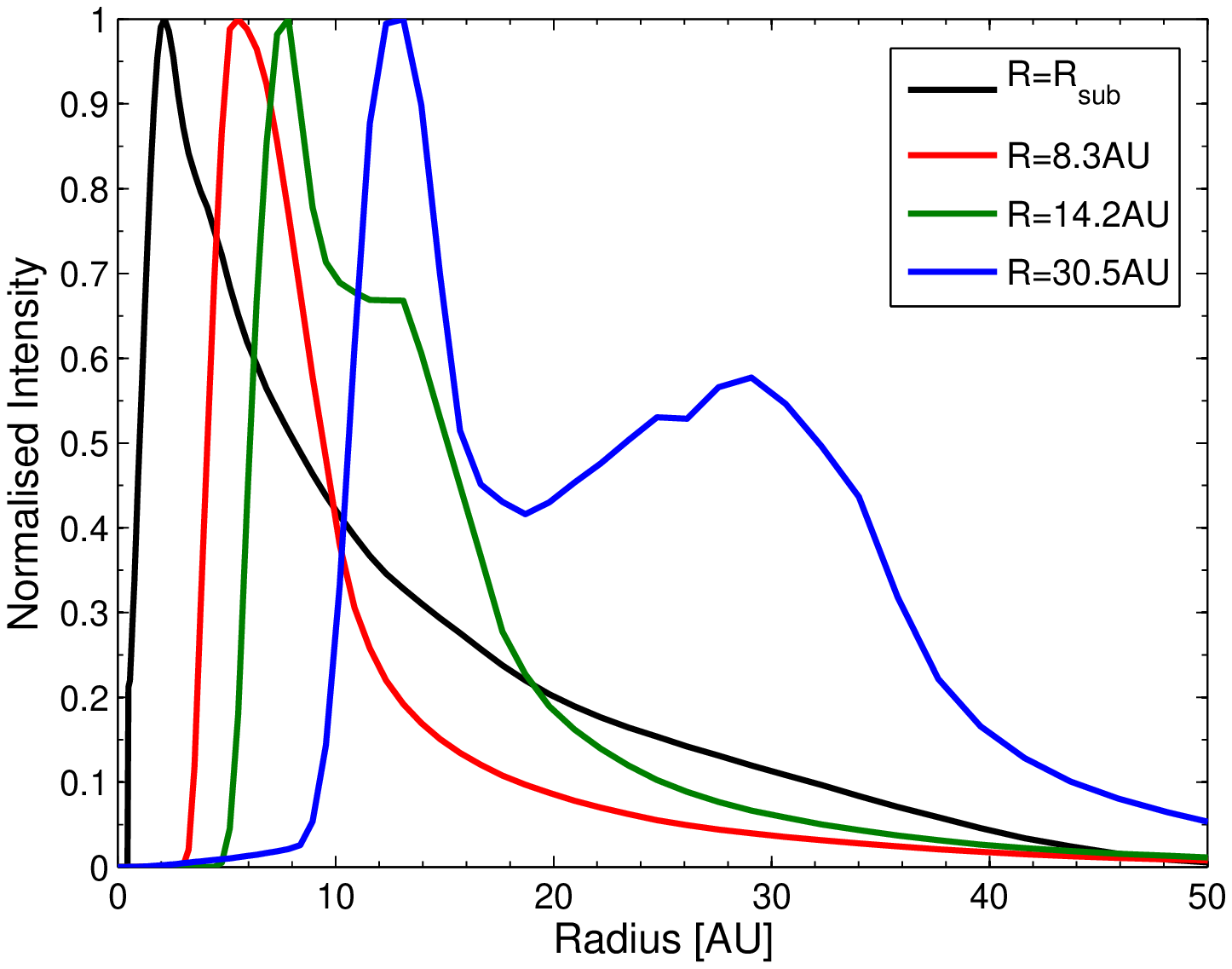}
\includegraphics[width=9.cm]{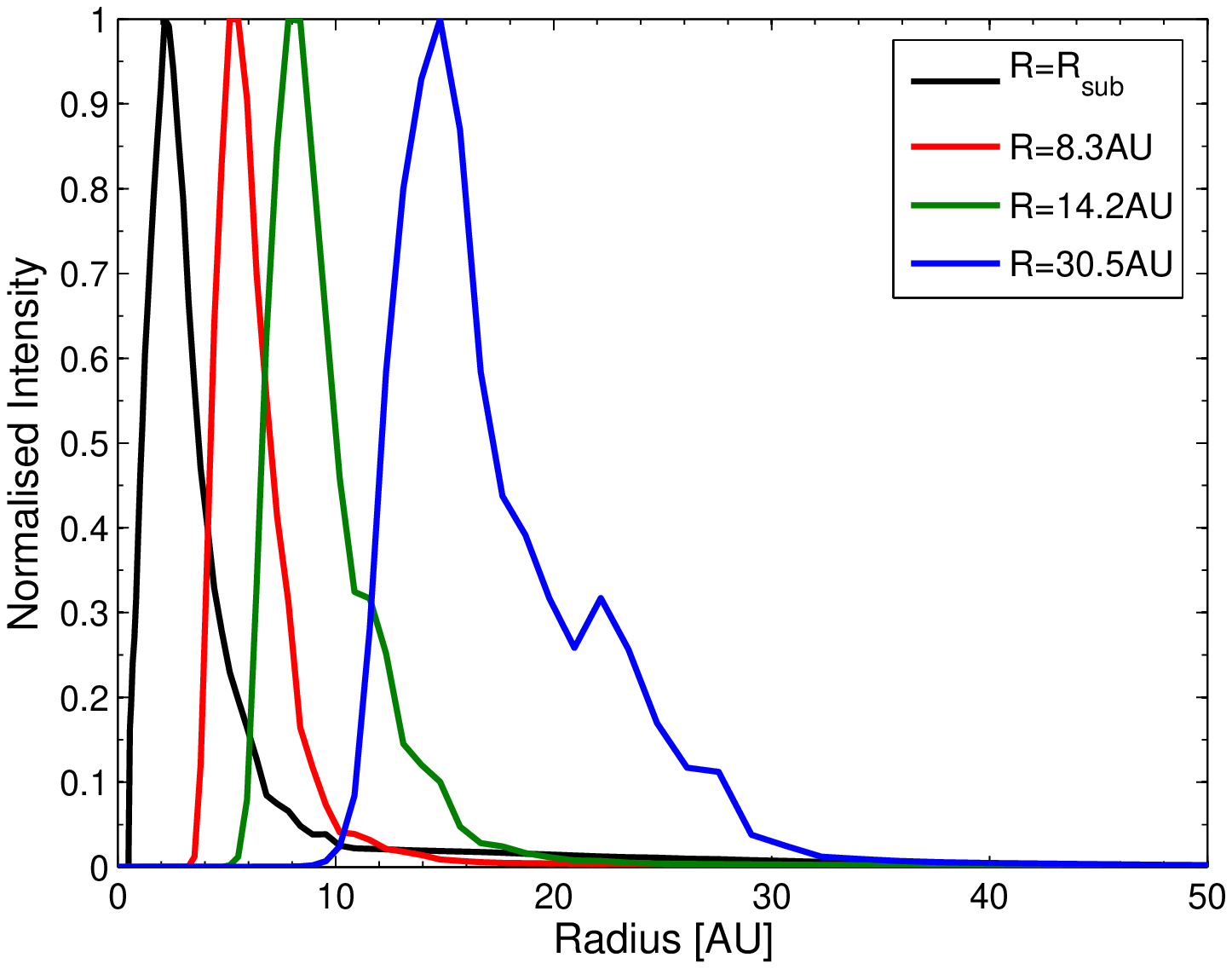}
\caption[]{Radial intensity profiles of NeII~12.8$\mu$m (left panels)
  and OI~6300{\AA} (right panel). The top panels show how the profiles
  respond to variations in the X-ray luminosity in the case of a
  primordial disc, while the bottom panels show the response to
  different hole sizes (R$_{in}$). R$_{sub}$ is the dust sublimation (destruction) radius. All models shown in the bottom panels are irradiated by an X-ray luminosity of 2$\times$10$^{30}$erg/sec.}
\end{minipage}
\label{fr}
\end{center}
\end{figure*}

The model's radial intensity profiles of the NeII~12.8$\mu$m and of the %added 's to model
OI~6300{\AA} line are shown in the left and right panels of
Figure~\ref{fr}. The top panels show the response of the radial
intensity profiles of these lines to changes in the X-ray
luminosity. Wider profiles are expected for lower X-ray luminosities
since these drive a wind that is less dense than in the higher X-ray
luminosity case, meaning that the emission region is more extended. 
The bottom panels of Figure~\ref{fr} show the radial intensity
profiles for discs irradiated by the same X-ray luminosity
(2$\times$10$^{30}$erg/sec), with the black line representing a
primordial disc and the red, green and blue lines representing discs
with inner holes of size 8.3, 14.2 and 30.5~AU, respectively. Both the
NeII and OI lines show a very clear sequence with inner hole size
suggesting that spectro-astrometry (e.g. Pontoppidan et al 2009) may
perhaps be used to test 
the evolutionary states of gaseous discs in the clearing phase. We draw
attention to the fact 
that the peak of the emission in all inner-hole cases is at radii
smaller than the dust inner radius of the disc. This is obvious by
looking at the structure of the wind and the location of the emission
region (see Figure~\ref{f3}). So the presence of gas in the inner hole
of these objects is simply due to gas that is being photoevaporated
from the inner edge of the disc.

\subsection{Caveats and limitations}

Before moving on to a comparison of our results with observations it
is worth stating a few caveats and limitations of our models. 

\begin{itemize}
\item It is important to realise that the current models can only
  provide a snapshot of the various flow solutions and line profiles
  that will arise. This means that it is currently not possible to
  present any form of 'interpolation' scheme for the line
  profiles and only general trends can be qualitatively
  extracted. There are two main reasons for this: (i) Most of the
  emitting regions (see Figure~\ref{f3}) is located in the subsonic
  region of the flow. In this region the flow velocity and direction
  is strongly and non-linearly affected by the interaction between the
  centrifugal force term and velocity divergence along the stream line
  (Ercolano et al. 2009a). Hence the variation of the blue shift with
  inclination and X-ray luminosity can only be understood in a
  qualitative way. The lack of a simple relation between line
  luminosity and inclination or X-ray luminosity is particularly
  evident in the behaviour of the OI line.
  (ii) The emitting region that dominates the luminosity for a given
  line can be considered a function of three variable: temperature,
  gas density and electron/ion fraction. All three influence the
  emissivity of a given pocket of gas and the effect they
  have on each other again arises from highly non-linear phenomena,
  meaning the emission region can change smoothly with X-ray
  luminosity (e.g. HI) or smoothly for some values of X-ray luminosity
  (e.g. L$_{X}>10^{29}$ erg/s) and rapidly for others
  (e.g. L$_{X}<10^{29}$ erg/s), as in the NeII line. 
\item The ionising spectrum, particularly in the soft X-ray and EUV
  range, is somewhat uncertain. Observations of this part of the
  spectrum are hindered by absorption from interstellar hydrogen along
  the line of sight or closer to the star. While we have attempted to
  use a realistic 
  spectrum (see Section~\ref{s:m} and discussion in Ercolano et al. 2009a),
  it is very likely that T-Tauri stars posses a range of spectra of
  different hardness. Accretion funnels close to the star may also
  screen the disc from EUV and soft X-ray effectively changing the
  hardness of the radiation field seen by the disc. The effects of
  screening on photoevaporation rates have been explored and discussed
  in detail by Ercolano et al (2009a). 
\item Our models assume azimuthal symmetry, however the effects of
  (possibly inclined) 
  magnetic fields on corotation radius may cause twists and warps at
  the disc's inner edge (e.g. Bouvier et al 2007;
  Romanova et al 2003; Long et al 2007). Furthermore accretion 
  funnels may only screen a fraction of the radiation resulting in
  azimuthally varying irradiation. As the structure of magnetic fields
  around T-Tauri stars is becoming clearer through observations
  and modeling (Donati et al 2007, 2008; Gregory et al 2008; Jardine
  2008; Hussain et al 2009; Skelly et al 2009), it will be very interesting to test 
  these effects on photoevaporation and disc emission.
\item The models presented in this paper are for a 0.7~M$_{\odot}$
  star. It is difficult to extrapolate these results to different mass
  stars, however models appropriate for different mass stars are
  currently being developed (Owen et al 2010b, in prep).
\item Our models do not account for molecular chemistry. While this is
  unlikely to produce a significant error in the photoevaporation
  rates and wind structure (see discussion in O10), it may affect the
  luminosities of low excitation temperature emission lines from
  neutral species (e.g. fine structure lines of OI and CI), which are
  produced in a region close to the atomic/molecular transition (the
  photodissociation region, PDR). We
  note that all the lines listed in Tables~\ref{t1} to~\ref{t4},
  however, are not significantly affected by the PDR, due to their high
  excitation temperatures (optical/UV collisionally excited lines)
  and/or ionisation requirements. 
\item Excitation rates of Ne$^+$ by collisions with neutrals are not
  known, therefore theoretical calculations of this line, including
  this work, only account for collisions with electrons. This is
  likely to result in the NeII~12.8$\mu$m line being underestimated by
  a factor of $\sim$2 (Glassgold et al 2007). 
\item The underlying disc flaring could be important since it
  determines the solid angle subtended by the disc to the ionising
  radiation. Owen et al (2010) briefly discuss this point with regards
  to mass loss rates in their Section 5, point vii. A discussion of
  the effects of flaring on line luminosities, specifically for the
  case of the NeII~12.8$\mu$m and NeIII~15.3$\mu$m lines, is given by
  Schisano, Ercolano \& G\"udel (2010). 
\end{itemize} 

\section{Discussion: The emission line spectra}

\begin{table*}
\caption[]{Line profiles from primordial discs irradiated by
  Log(L$_{X}$) = 28.3, 29.3 and 30.3 erg/s. The table lists the
  velocity of the peak, the full-width-half-maximum (FWHM) and the
  type of the profile (see Figure~4 for details), for ten inclinations from 0$^o$ to 90$^o$. The profiles were degraded to an instrumental resolution of R~=~100000. }
\label{t5}
\begin{tabular}{cccccccccc}
\hline
\multicolumn{10}{c}{NeII 12.8$\mu$m}\\
 & \multicolumn{3}{c}{Log(L$_{X}$)= 28.3 erg/s}& \multicolumn{3}{c}{Log(L$_{X}$)= 29.3 erg/s} & \multicolumn{3}{c}{Log(L$_{X}$)= 30.3 erg/s} \\
i$^o$ & v$_{peak}$ & FWHM   & Profile &v$_{peak}$ & FWHM & Profile & v$_{peak}$ & FWHM  & Profile  \\
      & [km/s]    & [km/s] &          &  [km/s]    & [km/s]&       &  [km/s]    & [km/s]         \\
\hline                               
0     &  0        &3.7  & B       &-0.25       &5.0 & B      &-2.25     & 10.2& A       \\
10    & -0.25     &6.0  & A       &-0.75       &10.8 & B     &-3.25     & 10.6& A       \\    
20    & -1        &10.1  & A      &-2.75       &13.9 & A     &-4.5      & 11.9& A       \\    
30    & -1.25     &13.6  & D      &-3.75       &15.4 & A     &-5.25     & 13.4& A       \\    
40    & -3/+1.75  &15.9  & C      &-4.25       &16.6 & A     &-5.5      & 15.0& A       \\    
50    & -3.5/+2   &17.6  & C      &-4.75       &17.6 & A     &-5.5      & 16.5& A       \\
60    & -3.75/+2.5&18.9  & C      &-2.25       &18.6 & D     &-5        & 17.8& A       \\
70    & -4/+3     &19.8  & C      &-4.75/+0.75 &19.4 & C     &-4.5      & 19.0& A       \\
80    & -4/+3.5   &20.4  & C      &-4.25/+1.75 &20.0 & C     &-3.75     & 19.8& A       \\
90    & $\pm$3.75 &20.8  & C      &$\pm$3.25   &20.6 & C     &$\pm$2.5  & 22.6& C       \\
\hline
%\multicolumn{10}{c}{NeIII 15.5$\mu$m}\\
% & \multicolumn{3}{c}{Log(L$_{X}$)= 28.3 erg/s}& \multicolumn{3}{c}{Log(L$_{X}$)= 29.3 erg/s} & \multicolumn{3}{c}{Log(L$_{X}$)= 30.3 erg/s} \\
%i$^o$ & v$_{peak}$ & FWHM   & Profile &v$_{peak}$ & FWHM & Profile & v$_{peak}$ & FWHM  & Profile  \\
%      & [km/s]    & [km/s] &          &  [km/s]    & [km/s]&       &  [km/s]    & [km/s]         \\
%\hline                               
%0     & -1.25     & 12.8 & A        &-3.75       &16.2 & A     & -4.25    &15.5 & A       \\
%10    & -1.5      & 13.9 & A        &-4          &16.6 & A     & -4.5     &15.8 & A       \\    
%20    & -2        & 16.0 & A        &-4.75       &17.4 & A     & -4.75    &16.6 & A       \\    
%30    & -2.75     & 18.0 & A        &-5.25       &18.4 & A     & -5       &17.8 & A       \\    
%40    & -3.25     & 20.0 & A        &-5.25       &19.6 & A     & -5       &19.2 & A       \\    
%50    & -3.25     & 21.9 & A        &-5          &20.7 & A     & -4.5     &20.7 & A       \\
%60    & -3        & 23.3 & A        &-4.25       &21.8 & A     & -3.75    &21.9 & A       \\
%70    & -2.5      & 24.5 & A        &-3.25       &22.7 & A     & -2.75    &22.9 & A       \\
%80    & -1.5      & 25.2 & A        &-1.75       &23.4 & A     & -1.5     &23.6 & A       \\
%90    & 0         & 25.5 & A        &0           &23.6 & A     & 0        &23.9 & A       \\
%\hline
\multicolumn{10}{c}{OI 6300~{\AA}}\\
 & \multicolumn{3}{c}{Log(L$_{X}$)= 28.3 erg/s}& \multicolumn{3}{c}{Log(L$_{X}$)= 29.3 erg/s} & \multicolumn{3}{c}{Log(L$_{X}$)= 30.3 erg/s} \\
i$^o$ & v$_{peak}$ & FWHM   & Profile &v$_{peak}$ & FWHM & Profile & v$_{peak}$ & FWHM  & Profile  \\
      & [km/s]    & [km/s] &          &  [km/s]    & [km/s]&       &  [km/s]    & [km/s]         \\
\hline                                
0     & -0.25     & 6.4  & A        &  -0.5      &10.0 & B     & -7.5     & 13.3& A       \\
10    & -4.75     & 14.0 & A        &  -3.75     &15.0 & B     & -7.75    & 13.0& A       \\    
20    & -5.75     & 15.9 & B        &  -6        &15.2 & B     & -8       & 12.6& A       \\    
30    & -5.5      & 16.0 & B        &  -7.25     &14.6 & B     & -8       & 12.6& A       \\    
40    & -5.5      & 16.2 & B        &  -8        &16.0 & B     & -7.75    & 13.1& A       \\    
50    & -5.25     & 17.0 & B        &  -7.5      &18.0 & B     & -7.25    & 13.9& A       \\
60    & -4.75     & 17.8 & B        &  -6.5      &20.1 & B     & -6       & 15.0& A       \\
70    & -4.25     & 18.5 & B        &  -5        &21.7 & B     & -4.5     & 16.0& A       \\
80    & -3.5      & 19.1 & B        &  -3.25     &22.7 & B     & -2.5     & 16.7& A       \\
90    & 0         & 19.5 & B        &  0         &23.1 & B     & 0        & 17.0& A       \\
\hline
\multicolumn{10}{c}{Hydrogen recombination lines}\\
 & \multicolumn{3}{c}{Log(L$_{X}$) = 28.3 erg/s}& \multicolumn{3}{c}{Log(L$_{X}$ = 29.3 erg/s} & \multicolumn{3}{c}{Log(L$_{X}$ = 30.3 erg/s} \\
i$^o$ & v$_{peak}$ & FWHM   & Profile &v$_{peak}$ & FWHM & Profile & v$_{peak}$ & FWHM  & Profile  \\
      & [km/s]    & [km/s] &        &  [km/s]    & [km/s]&       &  [km/s]    & [km/s]         \\
\hline
0     &0          & 7.2 &  A     &0           & 7.0&  A     &0          & 8.2&  A      \\
10    &0          & 8.2 &  A     &0           & 7.6&  A     &0          & 8.9&  A      \\    
20    &0          & 9.8 &  A     &0           & 9.2&  A     &0          & 10.1&  A      \\    
30    &0          & 11.5 &  B     &0          & 10.9&  B    &0          & 11.4&  B      \\    
40    &0          & 13.4 &  B     &0          & 12.8&  B    &0          & 12.9&  B      \\    
50    &-0.25      & 15.3 &  B     &0          & 14.6&  B    &-0.25      & 14.4&  B      \\
60    &-0.25      & 16.9 &  B     &-0.25      & 16.2&  B    &-0.25      & 15.7&  B      \\
70    &-0.5       & 18.1 &  B     &-0.5       & 17.5&  B    &-0.25      & 16.8&  B      \\
80    &-1         & 18.9 &  B     &0          & 18.2&  B    &-0.25      & 17.4&  B      \\
90    &$\pm$1     & 19.2 &  C     &$\pm$1     & 18.5&  C    &0          & 24.5&  B      \\
\hline
\end{tabular}
\end{table*}

\begin{table*}
\caption[]{Line profiles from inner-hole discs irradiated by
  Log(L$_{X}$ = 30.3 erg/s, with inner hole radii of 8.3, 14.2 and
  30.5~AU. The table lists the velocity of the peak, the
  full-width-half-maximum (FWHM) and the type of the profile (see Figure~4 for details), for ten inclinations from 0$^o$ to 90$^o$. The profiles were degraded to an instrumental resolution of R~=~100000. }
\label{t6}
\begin{tabular}{cccccccccc}
\hline
\multicolumn{10}{c}{NeII 12.8$\mu$m}\\
 & \multicolumn{3}{c}{R$_{in}$ = 8.3 AU}& \multicolumn{3}{c}{R$_{in}$ = 14.2 AU} & \multicolumn{3}{c}{R$_{in}$ = 30.5 AU} \\
i$^o$ & v$_{peak}$ & FWHM   & Profile &v$_{peak}$ & FWHM & Profile & v$_{peak}$ & FWHM  & Profile  \\
      & [km/s]    & [km/s] &          &  [km/s]    & [km/s]&       &  [km/s]    & [km/s]         \\
\hline                               
0     &   -1      & 15.0 & A        & -0.5       &13.3 & A     &  -0.5    &10.9 & A       \\
10    &   -1.25   & 15.4 & A        & -0.75      &13.9 & A     &  -0.5    &11.2 & A       \\    
20    &   -2      & 16.7 & A        & -1.25      &15.2 & A     &  -0.5    &12.1 & A       \\    
30    &   -2.75   & 19.7 & A        & -1.75      &17.6 & A     &  -0.75   &13.7 & A       \\    
40    &   -1.5    & 23.3 & D        & -1         &20.8 & D     &  -1.5    &15.9 & A       \\    
50    &   -1.5    & 26.2 & D        & -1         &23.5 & D     &  -0.5    &18.2 & D       \\
60    &   -1.5    & 28.4 & D        & -1.25      &25.8 & D     &  -0.75   &20.0 & D       \\
70    &   -7.25/+2.5  & 29.8 & C    & -5.75/+2.25 &27.2 & C     &  -3/+1.5 &21.3 & C       \\
80    &   -7/+4.25  & 30.2 & C      & -5.75/+3.75 &27.8 & C     &  -3.25/+2.25 &21.7 & C       \\
90    &   $\pm$5.75 & 30.1 & C      & -5/+5       &27.5 & C     &  $\pm$3.25 &21.3 & C       \\
\hline
%\multicolumn{10}{c}{NeIII 15.5$\mu$m}\\
% & \multicolumn{3}{c}{R$_{in}$ = 8.3 AU}& \multicolumn{3}{c}{R$_{in}$ = 14.2 AU} & \multicolumn{3}{c}{R$_{in}$ = 30.5 AU} \\
%i$^o$ & v$_{peak}$ & FWHM   & Profile &v$_{peak}$ & FWHM & Profile & v$_{peak}$ & FWHM  & Profile  \\
%      & [km/s]    & [km/s] &          &  [km/s]    & [km/s]&       &  [km/s]    & [km/s]         \\
%\hline                               
%0     & -0.5      & 17.6 & A        &-0.25       & 16.3& A     & 0        & 14.2& A       \\
%10    & -0.75     & 18.1 & A        &-0.25       & 16.9& A     & 0        & 14.4& A       \\    
%20    & -1.25     & 20.0 & A        &-0.5        & 18.5& A     & -0.25    & 15.2& A       \\    
%30    & -2.25     & 23.1 & A        &-1.25       & 21.3& A     & -0.25    & 16.5& A       \\    
%40    & -1.25     & 26.8 & D        &-2.25       & 24.7& A     & -0.25    & 18.1& A       \\    
%50    & -1.25     & 30.2 & D        &-1          & 28.1& D     & -0.5     & 19.9& A       \\
%60    & -1.25     & 33.0 & D        &-1.         & 30.9& D     & -0.25    & 21.5& A       \\
%70    & -7.5/+2.25& 34.9 & C        &-6.5/+2.0   & 32.9& C     & -0.25    & 22.6& A       \\
%80    & -7/+2.75  & 35.8 & C        &-6.5/+3.75  & 33.9& C     & -0.25    & 23.2& A       \\
%90    & $\pm$ .25 & 35.5 & C        &$\pm$5          & 33.7& C     & 0        & 23.5& A       \\
%\hline
\multicolumn{10}{c}{OI 6300~{\AA}}\\
 & \multicolumn{3}{c}{R$_{in}$ = 8.3 AU}& \multicolumn{3}{c}{R$_{in}$ = 14.2 AU} & \multicolumn{3}{c}{R$_{in}$ = 30.5 AU} \\
i$^o$ & v$_{peak}$ & FWHM   & Profile &v$_{peak}$ & FWHM & Profile & v$_{peak}$ & FWHM  & Profile  \\
      & [km/s]    & [km/s] &          &  [km/s]    & [km/s]&       &  [km/s]    & [km/s]         \\
\hline                                
0     & -0.25     & 17.1 & A        & 0          &14.7 & A     &  0       & 13.5& A       \\
10    & -0.25     & 17.2 & A        & 0          &15.0 & A     &  0       & 13.6& A       \\    
20    & -1        & 18.6 & A        & -0.25      &16.2 & A     &  0       & 14.2& A       \\    
30    & -2        & 22.4 & A        & -1         &19.1 & A     &  -0.25   & 15.7& A       \\    
40    & -0.75     & 26.9 & D        & -0.5       &23.3 & D     &  -0.75   & 18.5& A       \\    
50    & -9/+5     & 30.1 & C        & -6.5/+4.25 &26.6 & C     &  -0.25   & 21.5& D       \\
60    & -11.25/+7.25 & 32.4 & C     & -8.75/+6.25 &28.9 & C     &  -5.25/+4    & 23.8& C       \\
70    & -12.25/+8.75 & 34.0 & C     & -9.75/+7.75 &30.5 & C     &  -6.25/+5.25 & 25.3& C       \\
80    & -12/+10.25 & 34.9 & C       & -10/+8.5    &31.4 & C     &  -6.75/+5.75 & 26.2& C       \\
90    & $\pm$11.25    & 35.0 & C    & $\pm$9.5    &31.6 & C     &  $\pm$6.5   & 26.5& C       \\
\hline
\multicolumn{10}{c}{Hydrogen recombination lines}\\
 & \multicolumn{3}{c}{R$_{in}$ = 8.3 AU}& \multicolumn{3}{c}{R$_{in}$ = 14.2 AU} & \multicolumn{3}{c}{R$_{in}$ = 30.5 AU} \\
i$^o$ & v$_{peak}$ & FWHM   & Profile &v$_{peak}$ & FWHM & Profile & v$_{peak}$ & FWHM  & Profile  \\
      & [km/s]    & [km/s] &        &  [km/s]    & [km/s]&       &  [km/s]    & [km/s]         \\
\hline
0     &  0        &7.8  &  B     &    0       &7.1 &  B     &  -0.25    &6.5 &  B      \\
10    &  0        &8.3  &  B     &    0       &7.4 &  B     &  -0.25    &6.7 &  B      \\    
20    &  0        &9.9  &  B     &    0       &8.5 &  B     &  -0.25    &7.3 &  B      \\    
30    &  -0.25    &11.8  &  B     &    0       &10.0 &  B     &  -0.25    &8.1 &  B      \\    
40    &  -0.25    &13.6  &  B     &    0       &11.2 &  B     &  0        &8.8 &  B      \\    
50    &  -0.25    &15.1  &  B     &    0       &12.5 &  B     &  0        &9.6 &  B      \\
60    &  -0.25    &16.6  &  B     &    0       &13.8 &  B     &  0        &10.4 &  B      \\
70    &  -0.25    &18.1  &  B     &    -0.25   &15.0 &  B     &  0        &11.1 &  B      \\
80    &  -0.25    &19.0  &  B     &    -0.25   &15.7 &  B     &  0        &11.7 &  B      \\
90    &  0        &18.7  &  B     &    $\pm$0.75 &15.7 &  C     &  0      &11.8 &  B      \\
\hline
\end{tabular}
\end{table*}

The general behaviour of emission line intensities with X-ray
luminosity and inner hole size has already been discussion in
Section~\ref{s:a}. Here we further comment on an element-by-element basis
on the significance and observability of some of the transitions. 

\subsection{Carbon}
CI fine-structure lines are produced in a region that is very close to
the atomic to molecular transition. The luminosities calculated by
{\sc mocassin} for these transitions are likely to carry some error and have
therefore been omitted from the table and will be discussed in a
forthcoming paper where a simplified chemical network is
included. These lines may be interesting as probes of the gas at
larger radii (e.g. Ercolano et al 2009b). 

The CII fine structure line at 157$\mu$m, however, should be less
affected by the molecular/atomic transition and therefore we discuss
it further here. We report a luminosity of $\sim$10$^{-6}L_{\odot}$
for an X-ray luminosity of 2$\times$10$^{30}$erg/sec, this %removed source
should make this line an interesting target for {\it Herschel}
observations. We note that the luminosity we predict for this line is
almost three orders of magnitude larger than what was calculated by
Meijerink et al. (2008); as discussed in Ercolano et al (2008b), the
low critical density for this line (49~cm$^{-3}$) implies that it will be
preferentially emitted in low density regions, such as the upper disc
atmosphere and the flow, which were not included in the Meijerink
calculations. 

We also list a number of UV lines from CII, which are emitted
predominantly in the wind; the strongest of these transitions is the
doublet at 2327~{\AA}. 

\subsection{Oxygen}

\begin{figure}
\begin{center}
\includegraphics[width=9.cm]{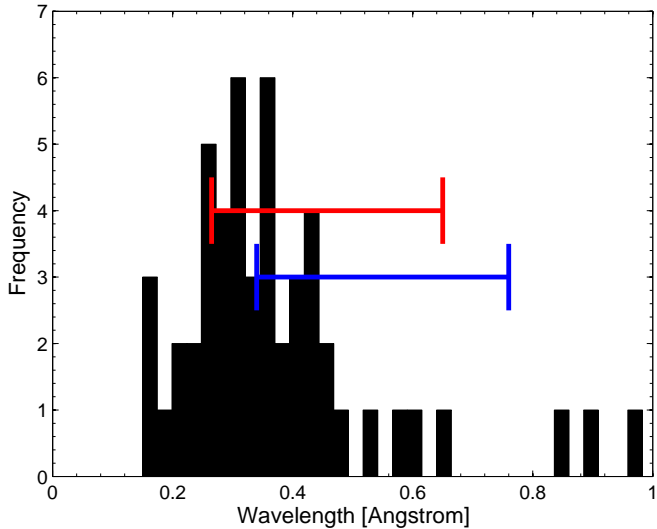}
\caption[]{OI 6300~{\AA} equivalent width histogram for the LVC sources in
  HEG95 compared to the model predictions for primordial discs (red
  line) irradiated by X-ray luminosities between 2$\times$10$^{28}$
  and 2$\times$10$^{30}$ erg/sec, and compared to inner hole sources
  (blue line) with hole sizes between $\sim$8 and $\sim$30~AU irradiated by an
  X-ray luminosity of 2$\times$10$^{30}$ erg/sec. The error in the
  equivalent widths is around the 10-20\% level.}
\label{f:ew}
\end{center}
\end{figure}

As in the case of carbon, we limit our discussion of neutral oxygen
emission to forbidden lines with high temperature Boltzmann terms and
exclude the OI fine-structure lines that may be uncertain in our
calculations for the reasons discussed above. 

The OI 6300{\AA} forbidden line deserves special attention. This line
was detected by HEG95 in the spectra of several
T-Tauri stars. Detection of this line in T-Tauri stars has been reported by various other authors including White \& Hillenbrand (2004) and Mohanty et al (2005).  The OI 6300{\AA} line profiles reported by HEG95 all
showed a double peak, with one component blue-shifted to a few hundred
km/s (high velocity component HVC) and a second component 
typically blue-shifted by only a few km/s (low velocity component,
LVC). The HVC
is generally attributed
to spatially unresolved emission from a dense jet close to the
star. The LVC, seen also in other forbidden lines, such as
OI~5577~{\AA}, and SII~6731~{\AA}, was interpreted by these authors as
being produced in a slow moving disc wind. This was evidenced 
by the fact that blue-shifts were largest for forbidden lines of lowest
critical density, suggesting that the low velocity component
accelerates away from the surface of the disc. HEG95 also noted 
that all optically thick discs that they observed had LVC forbidden line
emission detections. Furthermore, all the LVCs were blue-shifted to
within the measurements errors ($\pm$2 km/s), leading them to conclude
that, if the LVC arises from a disc wind, then all objects with
optically thick inner discs must have a disc wind.

While HEG95 did not ascertain the nature of the disc wind, a
photoevaporative wind origin was investigated by Font et
al (2004), who explored the ability of an EUV-driven photoevaporative wind
to explain the LVC lines observed in the HEG95 spectra. They obtained
encouraging results, which showed that 
forbidden line luminosities roughly of the observed magnitude can be
produced 
in the ionised wind from an EUV-photoevaporated disc, and highlighted
instead the failure of magnetic wind models in producing these lines. 
They identified, however, a discrepancy between their model and the
observations in the luminosity of the OI~6300~{\AA} line which
remained underpredicted by approximately two orders of magnitude in their models. 

The failure of the Font et al (2004) calculations to reproduce the
observed OI~6300{\AA} line luminosities results from the fact that an
EUV-driven wind is, by construction, fully ionised. The
OI~6300{\AA} line can only then be produced in a thin layer at the
bottom of the photoevaporating (ionised) layer. An X-ray driven photoevaporation
process, however, yields a disc wind that is predominantly neutral,
thus allowing abundant production of the OI~6300{\AA} line. Indeed we
predict luminosities of a few 10$^{-5}$~L$_{\odot}$ for
L$_{X}$~=~2$\times$10$^{30}$erg/s, roughly two orders of magnitude
higher than the 
value predicted by Font et al. (2004) and in the range of values
reported by HEG94. Figure~\ref{f:ew} shows a comparison of equivalent
widths observed in the LVC of OI~6300{\AA} by HEG95 (black histogram)
to the those predicted by our models (red and blue line). The red line
represents the primordial disc models irradiated by X-ray luminosities
between 2$\times$10$^{28}$ and 2$\times$10$^{30}$ erg/sec and the blue
line corresponds to inner hole sources with hole sizes between $\sim$8 and
$\sim$30~AU irradiated by an X-ray luminosity of 2$\times$10$^{30}$
erg/sec. The higher equivalent widths shown on the histogram 
are lines where even the LVC velocity range (-60 to 60 km~s$^{-1}$) is saturated by emission %added LVC range
from a jet (the x-axis is truncated in this plot and could go up to
7{\AA}). We note however that the range of hole sizes that we have
investigated is restricted, furthermore we do not have any inner-hole
models for lower luminosity.

The OI 6300{\AA} line is produced by collisional 
excitation, with neutral hydrogen being the dominant collider (rates
from Launay \& Roueff, 1977). We stress that in T-Tauri stars with
winds collisional excitation of the $^1$D$_2$ level of O will be the
dominant production mechanism of the OI~6300{\AA} line and not OH 
dissociation. UV photodissociation of OH, which may lead
to branching to the excited $^1$D$_2$ level of neutral O, producing
fluorescent emission of the OI 6300{\AA} and OI 6363{\AA} lines (St\"orzer \&
Hollenbach 2000) is generally ruled out for the HEG95 sample by the fact
that too large FUV fields are required and because of the difficulty
in reproducing the observed OI~6300~{\AA}~/~OI~5577~{\AA} ratio (see
also discussion in St\"orzer \&
Hollenbach 2000 and Font et al. 2004).

The luminosity of the OI~5577~{\AA} is underpredicted by our models
compared to the HEG95 observations. This is, however, not a cause of
concern, since this is most likely due to
the fact that collisions with neutral hydrogen (which are likely to
dominate as in the case of the OI~6300~{\AA} line) are not included in
our calculation, due to the unavailability of collision rates for the
$^1$S$_0$ level (Launay \& Roueff, 1977, only include the first four
levels, $^3$P$_2$, $^3$P$_1$, $^3$P$_0$ and $^1$D$_2$). The
OI~2973~{\AA} line is also a lower limit in our model for the same reason. 

In brief, an {\it X-ray driven wind model offers a natural route to
 the production of the observed LVC of the
OI~6300~{\AA} emission line}. In the next section we will discuss the
 profiles we obtained for this line and compare them with
 observations. 
 
Finally, OII and OIII forbidden lines are also listed in
Tables~\ref{t1} and \ref{t3}. OIII forbidden lines are not very strong
for primordial disc models and therefore are only listed for
inner-hole models. Inner hole model winds 
are less dense and hotter than those of primordial discs, hence
favouring the production of nebular-like lines like the OII and OIII
listed in the table. We draw particular attention to the OIII
88.3$\mu$m line, which has luminosities of around 10$^{-7}$L$_{\odot}$
for inner hole sources, and may be just detectable with PACS for the
nearer sources. 

\subsection{Neon} 

Since the prediction by Glassgold et al (2007) and the first
detections of the NeII~12.8~$\mu$m line in T-Tauri discs (Pascucci et
al 2007; Espaillat et al 2007; Lahuis et al 2007; Herczeg et al 2007),
this line has    
received a lot of attention in the literature, both observationally
(G\"udel et al 2010, submitted; Pascucci \& Sterzik 2009; Najita et al
2009; Pascucci et al 2009; van Boekel et al 2009; Flaccomio et al
2009) and theoretically (Meijerink et al, 2008; Schisano et al. 2009; 
Alexander 2008; Hollenbach \& Gorti 2009). This line was originally
proposed as a new probe of X-ray irradiation of T-Tauri
discs, but a clean correlation with X-ray luminosity does not seem to
exist even in the latest data. The large scatter
in the L(NeII) for stars of a given L$_X$ may be due to differences in
the irradiating spectrum or the disc structure (Schisano et al 2009),
variability, or, for the higher L(NeII) cases, to stellar jets
(Hollenbach \& Gorti 2009; Shang et al 2010, submitted).  

The L(NeII) we calculate from our models are in the range of
those observed and show a roughly linear dependance with X-ray
luminosity (see also Section~\ref{s:a}). What our models show, however, is
that this line is produced almost exclusively in the photoevaporating
wind, as clear from the blue-shifted profile for non-edge on
inclinations (Table~\ref{t5} and Figure~\ref{f1}). This
opens the exciting prospect of using this line as a probe of
photoevaporative winds as we will discuss further in the next
Section. It is also clear that a clean linear relationship with L$_X$
is not expected (or found) given that the wind structure and the
NeII emitting region change in response to different X-ray
luminosities. In this perspective it is easy to understand the scatter
in the L(NeII)-L$_X$ relation noticed by G\"udel et al (2010, submitted) even for
the non-jet sources.

L(NeII) is about a factor of two larger for inner-hole models than a
primordial model of the same X-ray luminosity and this is simply due to
the larger emission region of the inner-hole sources (see
Figure~\ref{f3}). We note that while Schisano et al. (2009) used disc-only
models to probe the dispersion in the 
L(Ne)-L$_X$ 
relationship, many of their conclusions with regards to the nature of
the scatter (e.g. hardness of the ionising radiation field) are equally
applicable to a line that is produced in the 
wind. 

The NeIII to NeII ratio for our models is never lower than $\sim$0.1,
while observations generally indicate lower values, e.g. $\sim$0.06
(Lahuis et al 2007), $\sim$0.045 (Najita et al 2010), although in the
case of WL5 a value of $\sim$0.25 is reported by Flaccomio et al (2009).
 Unfortunately only very few detections of this line are currently
 available and it is therefore hard to draw statistically significant
 conclusions. To make matters worse this ratio is quite uncertain in
 theoretical models due to the lack of neutral collision rates, as
 mentioned in Section~\ref{s:a}, and uncertainties in the charge
 exchange coefficients. In view of the current observational and
 theoretical difficulties with this line ratio, its diagnostic power
 cannot yet be harnessed. 

\subsection{Magnesium}
We predict a rather strong MgI line at 4573~{\AA}
($\sim$10$^{-5}$L$_{\odot}$ for
L$_{X}$~=~2$\times$10$^{30}$erg/sec). This line has been detected to
date, in 
three young low mass stars, namely VY Tauri (M0; Herbig 1990), XZ
Tau B (M1; White \& Ghez 2001) and very recently in the low-mass
variable TWA~30 (Looper et al. 2010, submitted). Detection of this line in
M-stars is rendered possible by the much lower stellar continuum in the
B-band compared to higher mass stars. The strength of this line
implies that it could be an important coolant and should 
 be considered when thermal calculation are carried out. 

\subsection{Silicon}
We include Si lines in Tables~\ref{t1} and \ref{t3} and draw the reader's
attention to the SiII~34.8$\mu$m transition which has a luminosity of
approximately 10$^{-6}$L$_{\odot}$ for L$_{X}$~=~10$^{30}$erg/sec. We
caution however that Si lines and Mg lines are strongly affected by
depletion onto dust grains which is very poorly constrained. While a %changed an to a
hindrance for the prediction of accurate line luminosities, the
dependance of this lines on depletion factors may actually be used to
obtain clues about the disc evolution through comparison of models with %change latter to dust evolutoion through
observations. 

\subsection{Sulfur}
We predict a number of SII, SIII and SIV lines all with fairly large
luminosities. In particular our models predict SII 6717{\AA} and
6731{\AA} lines in the range observed by HEG95\footnote{Note that here
  we only compare with HEG95 for the LVC detections}. The ratio of the
6731{\AA} to the 6717{\AA}, which is a useful density diagnostic,
is approximately two in our models, and it compares well with the
value reported for most of the sources in HEG95. A ratio of two at
$\sim$7000~K corresponds to an {\it electron} density of just below
10$^4$cm$^{-3}$, again consistent with a wind origin of this
line. Furthermore, all of the detections for the LVCs of these lines in
HEG95 show a small blue-shift, reinforcing the conclusion that all
discs show evidence of a wind which can be produced by photoevaporation. 
The OI to SII ratios also compare favorably, when the OI line luminosites observed are in the range 
predicted by the model (i.e. $<10^{-4}$L$_\odot$), however we note that this ratio can be as large a one order of magnitude for the higher 
OI luminosities detected by HEG95, these detections correspond to objects where the jet emission has saturated the LVC and are associated with large equivalent widths in the LVC OI line (e.g. CW Tau, DQ Tau and HN Tau).

\subsection{Hydrogen}

Tables~\ref{t2} and \ref{t4} show the luminosities of the
hydrogen recombination lines. The HI(7-6) and HI(6-5) transition
were detected by Pascucci et al. (2007), Ratzka et al. (2007), and
a larger set of IR lines have been detected from the transition object TW Hya by 
Najita et al. (2010). The reported luminosities of those
lines are of the same magnitude as for the NeII~12.8$\mu$m line also detected in the 
same objects. Our models, on the other hand, can only produce hydrogen
recombination lines that are almost two orders of magnitude less
luminous than the NeII~12.8$\mu$m. It seems therefore unlikely that a
disc or wind origin can explain the reported hydrogen recombination
line luminosities. This had already been noticed by Hollenbach \&
Gorti (2008), who analytically derive L(HI 7-6)/L(NeII)$\sim$0.008,
similar to what we find (0.015). We agree with Hollenbach \& Gorti's
conclusion that these IR hydrogen lines must originate in a 
region of very dense plasma (where NeII~12.8$\mu$m would be
collisionally quenched) close in to the star, possibly in an
accretion shock. This is in line with earlier works (e.g. Hartmann et
al 2004; Muzerolle et al 1998, 2000; Kurosawa et al 2005) that H
recombination lines mostly originate in the magnetospheric accretion
flow. 

\section{Discussion: Blue-shifted line profiles, a smoking gun for photoevaporation}

\begin{figure}
\begin{center}
\includegraphics[width=8.cm]{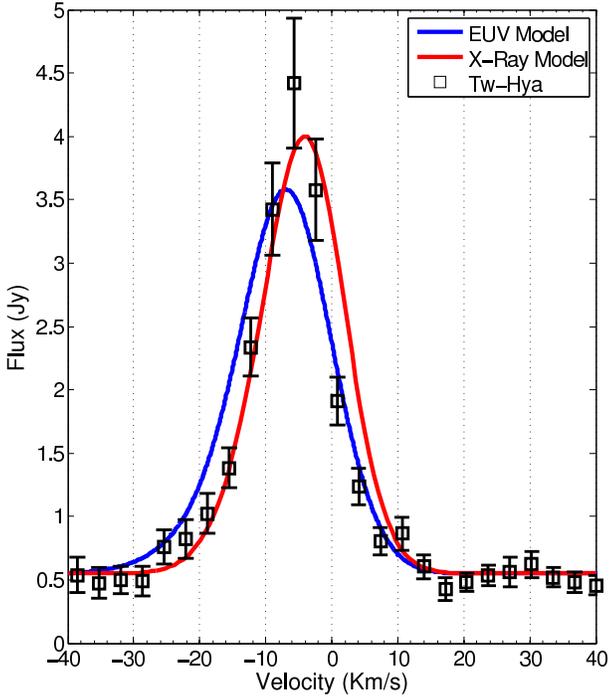}
\caption[]{Profile of the NeII~12.8$\mu$m fine-structure line as
  predicted by the X-ray driven photoevaporation model of O10 --blue line--, the EUV-only model
  of Font et al. 2004 (see Alexander 2008) --red line--, compared to the
  observation of TW-Hya by Pascucci et al. (2009) --black
  squares--.}
\label{f5}
\end{center}
\end{figure}

The recent observation of a blue-shifted NeII~12.8$\mu$m line profile
by Pascucci et al. (2009, P09) in a few well known YSOs with discs,
including TW~Hya, has been interpreted as evidence for
photoevaporation. As shown by Alexander (2008) the EUV-wind solution
of Font 
et al. (2004) produces a good match to the data (red line in
Figure~\ref{f5}), as does the X-ray driven wind of O10 discussed in this
paper (blue line in Figure~\ref{f5}). The X-ray wind model used for
this figure is that of the primordial disc model irradiated by an
X-ray luminosity, Log(L$_X$)~=~30.2~erg/sec. We stress that neither
the EUV-wind or X-ray wind models were {\it statistically fit}
to the data to 
produce the match. In both cases a model was chosen from the set of
existing ones that best reproduced the observations. Both the EUV wind
and the X-ray wind models initial parameters could be tweaked to
produce a perfect fit to the data, but this was not attempted in this
paper or by Alexander (2008) as the point of Figure~\ref{f5} is only to
show that profiles in the range of those observed can be reproduced by
photoevaporation models. 
 
As both the EUV- and X-ray model produce a good match to the NeII data, it 
seems that this 
particular diagnostic cannot be 
used to discriminate between them. The OI~6300{\AA} line intensities
and LVC profiles, on the other hand, can only be reproduced by an
X-ray driven photoevaporation model, which predicts a largely neutral
wind. 

While a blue-shifted emission line profile is a clear tell-tale sign
of outflows, the detection of non-blue-shifted lines (e.g. Najita et
al. 2009) cannot be unambiguously interpreted as absence of a wind. It
is clear from Figure~\ref{f1} and~\ref{f2} and the discussion in the
previous section, that there are many reasons why a line like
NeII~12.8$\mu$m or OI~6300{\AA}, that has a significant blue-shift in
some cases, may instead appear centred in others --e.g. inclination,
inner hole size, X-ray luminosity--, also different profiles respond
differently to the change in these parameters, such that the
relative centroid of, say, the NeII~12.8$\mu$m with respect to the
OI~6300{\AA} line may also shift with the shift in X-ray luminosity
(see Figure~1). 

It should also be made clear that photoevaporation models predict that
a large number of primordial discs, not only so-called transition
discs, should have a disc wind and therefore produce blue-shifted
emission lines. This is in agreement with HEG95's 
observations that showed blue-shifted LVC of OI~6300{\AA} for all the
YSOs with discs. The detection of blue-shifted lines does not indicate
that a given disc is evolved, all that it indicates is that energetic
radiation from the central star is able to reach the disc establishing
a flow. Our models predict intensities and blueshifts for this line
that are in agreement with observations, suggesting that the
LVC of OI~6300{\AA} is indeed a good candidate for tracing
photoevaporating disc gas. Further spectrally resolved observations of
this line would be extremely helpful in testing the presented model. %added in testing the presented model 

Pascucci et al. (2009) found that of the seven sources they
observed with VISIR, a NeII detection was only obtained in the
'transition' (i.e. dust inner hole) sources. They interpreted 
their result as 
evidence of EUV-driven photoevaporation\footnote{At the time of
  publication of Pascucci et al. (2009) line profiles had only been
  calculated for the EUV-driven wind model
  (Font et al. 2004, Alexander 2008).} and concluded that this can
occur only at a late stage in the disc evolution. 
However we note that the detection limits at the distance of the
primordial disc sources which showed no evidence of NeII emission
are of a few 10$^{-6}$L$_{\odot}$, these are quite high and
would only allow the strongest NeII sources to be detected. Indeed
G\"udel et al (2010, submitted) show that, excluding sources with known jets, many
cTTS have L(NeII) lower than a few 10$^{-6}$L$_{\odot}$. The NeII
luminosities predicted by our models for a primordial 
disc irradiated by Log(L$_X$)=30.2 would be barely detectable at the
sensitivity of the Pascucci et al. (2009) observations. 
The fact that NeII~12.8$\mu$m was only detected in three of the
sources of the P09 sample and that these happened to be transition
discs may just be a coincidence, or, more likely, it may be due to the
fact that, as
predicted by our models, the NeII line is brighter for transition
discs. The latter explanation however would only stand if the inner
hole of these sources were devoid of gas (or at least the gas was %changed were to was
optically thin to both X-rays and the interested transitions). However
all of these 'transition' discs are still accreting and depending on
the gas densities in the inner hole, perhaps a primordial disc model
may be more appropriate. 

One should also point out that the magnitude
of the blue-shift depends entirely on the size and location of the
emitting region. The blueshift will be largest for inclinations where
the emitting region is such to maximise the line-of-sight component of
the velocity vectors. The blueshift of a line is therefore not
necessarily maximised for a completely face on disc. A good example of
this is provided by the OI~6300{\AA} profile of the Log(L$_X$)~=~29.3
erg/sec primordial disc model.

While the current observational and theoretical emission line profile
sample is still too small to draw any significant conclusions, these
observations highlight the potential of using high resolution
spectroscopy of YSOs and detailed profile modelling to learn about the
dispersal mechanism of protoplanetary discs.

\section{Conclusions}

X-rays have recently been shown to drive powerful outflows (Ercolano
et al. 2008b, 2009a, Owen et al 2010a). The
recent radiation-hydrodynamic 
calculation of Owen et al. (2010a) have demonstrated that that the two
time-scale dispersal of protoplanetary discs around solar-type stars
is successfully reproduced by this model. Further evidence is
provided by the observation of shorter disc lifetimes in
low-metallicity environments (Yasui et al. 2009), a behaviour that is
expected if X-ray driven photoevaporation dominates the dispersal of
discs and argues against planet formation as the main dispersal
mechanism (Ercolano \&  Clarke, 2009). Furthermore the X-ray
spectra of T-Tauri stars are fairly well characterised (e.g. G\"udel
et al 2007) and
the observed luminosities imply that significant X-ray photoevaporation
must occur, unless the inner source is
completely obscured due to accretion funnels thicker than
10$^{22}$cm$^{-2}$ (Ercolano et al. 2009a) and covering all lines of sight
from the source to the disc.

Encouraged by these results, in this paper we have presented an
atlas of emission lines from atomic and low-ionisation species emitted
by photoevaporating protoplanetary discs around solar type stars. We
have used disc and wind density distributions obtained by the
radiation-hydrodynamic calculations of Owen et al (2010a) in the X-ray
photoevaporation paradigm and have considered discs irradiated by
a range of X-ray luminosities and at various stages of clearing,
from primordial (i.e. extending all the way into the dust destruction
radius) to inner hole sources with various hole sizes. Line profiles
for the NeII~12.8$\mu$m, OI~6300{\AA} and HI
recombination lines have also been calculated and should provide
guidance for the interpretation of high-resolution spectroscopy of
YSOs. 

We compare the predictions of our models to available data in the
literature and find agreement for the line intensities and
profiles. We draw particular attention to the good agreement with the
observations of Hartigan et al (1995) of the OI~6300{\AA} line intensity
and profile in the spectra of T-Tauri stars. This had been a problem
for previous, EUV-driven photoevaporation models (e.g. Font et
al. 2004) which underestimated its luminosity by a couple
of orders of magnitude. The problem with these previous models lay in
the fact that EUV winds are, by construction, ionised, while X-ray
photoevaporation provides a mechanism to drive warm neutral winds, a
condition necessary for the production of blue-shifted collisionally
excited OI~6300{\AA} and OI~5577{\AA}. 

As noted by Font et al (2004), magnetic wind models systematically
underpredict the luminosities of forbidden lines and cannot fit the
observed profiles (e.g. Garcia et al 2001). The main problem with
magnetic wind models is that they are too cold -- an X-ray
driven photoevaporative wind naturally produces temperature,
ionisation degrees and consistent kinematic patterns able to produce
forbidden line intensities and profiles in agreement with the
observations. These results suggest that, while magnetic disc winds
may work in conjunction with photoevaporation, they are unlikely to be
operating alone.

We analyse our results for the NeII~12.8$\mu$m line profiles in
the context of recent high resolution observations of Pascucci et al
(2009) and Najita et al. (2009). We discuss how the detection of a
blue-shifted line is a clear sign of an outflow, but caution against
interpreting non-blue-shifted profiles or non-detections as absence of
a disc wind. We argue that the non-detections for the primordial discs
in the Pascucci et al (2009) sample could be explained by insufficient
sensitivity of the observations that are only able to probe the upper
end of the NeII luminosity distribution. The non-blue-shifted line
centroids in the Najita et al. (2009) sources are probably due to
unfavourable (high inclination) viewing angles or to the effect of
seeing blue and redshifted emitting material through a large inner
hole, as may be the case of GM~Aur, although we caution that this
explanation is only valid if the gas, known to be present in the inner
(dust) hole of this object, is optically thin to X-ray radiation
and to the line in question. 

Finally, our models cannot reproduce the observed emission line
luminosities of IR hydrogen recombination lines, which for some
objects are reported to be as bright as the NeII~12.8$\mu$m
(Pascucci et al 2007; Ratzka et al 2007). As suggested by Hollenbach \& Gorti
(2009), these lines are 
probably produced in a dense, ionised region that is very close to the
central star. High resolution spectroscopy of these lines is needed to
further constrain their origin. 

\section*{Acknowledgments}
We thank the anonimous referee for an extremely detailed report that
helped improve the clarity of our paper. 
We thank Ilaria Pascucci, Manuel G\"udel and Joan Najita for helpful
discussion and comparison of the observational results with our
models. We thank Cathie Clarke, Al Glassgold, Greg Herczeg and
Richard Alexander for helpful comments and a critical read of the
paper. We also thank Richard Alexander for providing the NeII line
profile from the EUV-photoevaporation model. We thank Eugenio Schisano
for helping with the line profiles. BE is supported by a Science and
Technology Facility Council Advanced Fellowship. JEO acknowledges support of a STFC PhD studentship.  %added me to acknowledgements
This work was performed using the Darwin Supercomputer
of the University of Cambridge High Performance Computing Service
(http://www.hpc.cam.ac.uk/), provided by Dell Inc. using Strategic
Research Infrastructure Funding from the Higher Education Funding
Council for England.

\end{document}